\documentclass[useAMS, usenatbib]{mn2e}
\usepackage{graphicx}  

\title[Spiral galaxies in cosmological simulations]{Towards a more realistic population of bright spiral galaxies in cosmological simulations}
\author[M. Aumer et al.]
{Michael Aumer$^{1,2}$ \thanks{E-mail:maumer@mpa-garching.mpg.de (MA)}, Simon D.M. White$^{1}$, Thorsten Naab$^{1}$, Cecilia Scannapieco$^{3}$\\
$^{1}$Max-Planck-Institut f\"ur Astrophysik, Karl-Schwarzschild-Str. 1, 85748 Garching, Germany\\
$^{2}$Excellence Cluster Universe, Boltzmannstr. 2, 85748 Garching, Germany\\
$^{3}$Leibniz-Institut f\"ur Astrophysik Potsdam (AIP), An der Sternwarte 16, 14482 Potsdam, Germany}

\begin{document}

\date{Accepted 2013 July 03. Received 2013 July 02; in original form 2013 March 28}

\pagerange{\pageref{firstpage}--\pageref{lastpage}} \pubyear{2013}

\maketitle

\label{firstpage}

\begin{abstract}
We present an update to the multiphase SPH galaxy formation code by Scannapieco et al.
We include a more elaborate treatment of the production of metals, cooling
rates based on individual element abundances, and a scheme for the turbulent diffusion
of metals. Our SN feedback model now transfers energy to the ISM in kinetic and thermal form,
and we include a prescription for the effects of radiation pressure from massive
young stars on the ISM. We calibrate our new code on the well studied Aquarius haloes and then
use it to simulate a sample of 16 galaxies with halo masses between $1 \times 10^{11}$ and $3 \times 10^{12} M_{\odot}$.
In general, the stellar masses of the sample agree well with the stellar
mass to halo mass relation inferred from abundance matching techniques for redshifts $z=0-4$.
There is however a tendency to overproduce stars at $z>4$ and to underproduce them at $z<0.5$
in the least massive haloes. Overly high SFRs at $z<1$ for the most massive haloes 
are likely connected to the lack of AGN feedback in our model.
The simulated sample also shows reasonable agreement with observed star formation rates,
sizes, gas fractions and gas-phase metallicities at $z=0-3$. Remaining discrepancies can be connected
to deviations from predictions for star formation histories from abundance matching. At $z=0$, the model galaxies show realistic
morphologies, stellar surface density profiles, circular velocity curves and stellar metallicities,
but overly flat metallicity gradients. 15 out of 16 of our galaxies
contain disk components with kinematic disk fraction ranging between 15 and 65 $\%$.
The disk fraction depends on the time of the last destructive merger or
misaligned infall event. Considering the remaining shortcomings of our simulations
we conclude that even higher kinematic disk fractions may be possible
for $\Lambda$CDM haloes with quiet merger histories, such
as the Aquarius haloes.
\end{abstract}


\section{Introduction}

In the standard paradigm of cosmic structure formation, galaxies
form through cooling and condensation of gas within dark matter haloes \citep{whiterees, fall}. 
Collisionless $N$-body simulations of the dark matter component have been able to reproduce
the observed large-scale structure of the universe with high accuracy 
in the $\Lambda$ Cold Dark Matter ($\Lambda$CDM) version of this paradigm \citep{millenium}.
Semi-analytic models relying on these simulations and simple analytical prescriptions for the baryonic
component are capable of reproducing the detailed systematics of the galaxy population \citep{guo2}. 
To properly understand the complex dynamical interactions between gas, stars and dark matter in galaxies
requires however cosmological hydrodynamical simulations.

The complexity of properly modeling the many baryonic astrophysical processes which play a role in the formation
of galaxies has led to a long list of models over the last two decades (e.g. \citealp{navarro94, navarro2,
abadi, governato, cs09, agertz}). Despite significant recent progress (e.g. \citealp{sales, eris, 4disk}), these
simulations continue to be plagued by a range of problems, with different codes often producing very different galaxies
for the same initial conditions \citep{okamoto5, keres, aquila}. 

The formation of realistic disc galaxies has been shown to be especially problematic. The cooling and condensation
of too much low-angular momentum gas (overcooling) and the loss of angular momentum from baryons to the dark matter
have been a problem since the first efforts to simulate cosmological galaxy formation \citep{navarrobenz}.
Once disks have formed they have been shown to be susceptible to destruction by major mergers \citep{toomre},
by the infall of satellite galaxies \citep{toth} and by accretion of misaligned gas \citep{cs09}(CS09 hereafter). Moreover, the reorientation
of disks \citep{aw, okamoto} and disk instabilities (e.g. \citealp{noguchi, diplom}) can enhance the bulge-to-disk
ratio.

In addition to these problems, stellar mass to halo mass relations from abundance matching techniques have shown that the
majority of simulations over-produce stars \citep{guo, sawala}, especially at high $z$ (CS09, \citealp{moster}).

Modeling the injection of energy from supernova explosions into the surrounding gas (e.g. \citealp{cs06, stinson})
has been shown to be a possible mechanism to prevent overly efficient gas cooling, for driving large-scale outflows,
for removing low angular momentum material and thus for producing more realistic disk galaxies (e.g. \citealp{cs08, brook}).
Interestingly, simulations applying rather weak feedback have been shown to be more successful 
in reproducing early-type galaxies \citep{naab, oser, peter}.
Apart from supernovae, AGN (e.g. \citealp{springel05}) and cosmic rays \citep{uhlig} have been considered
as possible sources of feedback. 

Simulations including empirical models of momentum-driven winds have been shown to improve
the agreement with observed properties of the galaxy population and the intergalactic medium \citep{od06,od10}.
Recently, the input of momentum and energy from massive young stars 
in form of stellar winds and radiation pressure prior to their explosion as SNe 
has been studied in more detail. \citet{hopkins} showed that momentum input into the ISM
from radiation pressure may play a key role in regulating star formation. \citet{stinson13} included thermal feedback from young stars
and thus were able to achieve significantly better agreement of star formation histories with observations.
\citet{agertz13} presented the most complete model so far of mass, momentum and energy feedback from stars during all stages
of their evolution, and concluded that radiation pressure from young stars has the strongest effect on the ISM.

Apart from feedback, a more realistic modeling of star formation and the ISM has been shown to help
in making simulated galaxies with more realistic properties. \citet{governato10} found that
increasing the threshold density used in the modeling of star formation to more realistic values
leads to more concentrated star formation, more efficient feedback energy input into the ISM, and 
the formation of galaxies with higher disk fraction (see also \citealp{eris}).
\citet{christensen} have shown that the implementation of a model for the formation of molecular hydrogen
(see \citealp{gnedin}) can amplify the effect of higher threshold densities.

In \citet{aw} (AW13 hereafter), we have shown that it is possible
to form a realistic disk galaxy within a triaxial, substructured and growing
CDM halo, as long as the disk forms predominantly at late times ($z<1$), the angular momentum vector of the infalling gas
has a roughly constant and appropriate orientation and there are no major mergers or other significant changes in the configuration
of the halo. In the \textit{Aquila Code Comparison Project} (\citealp{aquila}, CS12 hereafter) one of the haloes considered in AW13
was simulated with 15 different galaxy formation codes yielding 15 model galaxies with widely varying properties.
Although several models compared well to a subset of the observations considered, none of the models yielded a fully realistic disk galaxy.
In this paper, we intend to show how modifications to the models of CS09, motivated by the findings of CS12 and some
of the recently discussed solutions to problems discussed above, can lead to the formation of significantly more realistic
disk galaxies. We also use our model to study a sample of haloes, which has been previously studied by \citet{oser}
in the context of the formation of massive, early-type galaxies. The combined sample comprises haloes with very quiet merger histories
as well as low-$z$ mergers, and is thus well suited to be compared to a range of objects in the observed galaxy population.

Our paper is organized as follows:
In Section 2 we describe the updates we applied to our code.
In Section 3 we introduce the sample of initial conditions we use.
In Section 4 we discuss the star formation histories of our galaxies.
In Section 5 we analyze their kinematical and structural properties.
In Section 6 we compare our sample to observed scaling relations.
Finally, we conclude in Section 7.

\section{The code}

For our simulations, we use the TREESPH code GADGET-3, last described in \citet{gadget}.
\cite{cs05,cs06} introduced models for stellar metal production, metal line cooling, star formation, SN 
feedback and a multiphase gas treatment to be used with GADGET-3, which we use as the basis
for our code update.

\subsection{Multiphase model and star formation}

We begin with a description of the parts of the model, which have remained unchanged.
The code is unique in decoupling SPH particles with very different thermodynamic properties by preventing
them from being SPH neighbours. Particle $i$ decouples from particle $j$ if
\begin{equation}
A_{i} > 50.\times A_{j} \,\,\,\,\,\,\,\,\,\,\,\, \rm {and} \,\,\,\,\,\,\,\,\,\,\,\, \mu_{ij} < c_{ij}.
\end{equation}
Here $A$ is the entropic function \citep{sh02}, $c_{ij}$ is the pair-averaged sound-speed and $\mu_{ij}$ is
the relative velocity of the particles projected onto their separation vector. \citet{marri} showed that
the second condition is needed to avoid decoupling in shock waves, which can lead to unphysical effects.
This multiphase gas treatment has been implemented to make a realistic co-existence of hot and cold phase gas
(as observed in the ISM) possible in SPH. It has also been shown to allow a more realistic modeling of
energy and metal injection from stars to the distinct components of the ISM.

To model the formation of stars, we assume that gas particles are eligible for this process, if their density
is above a threshold density $n_{\rm{th}}$. Whereas CS09 used $n_{\rm{th}}\sim 0.03 \;\rm{cm}^{-3}$,
\citet{governato10} have argued that significantly higher threshold densities are needed to form
realistic galaxies. While we can confirm their conclusions, we caution that the effect
of varying this parameter depends significantly on the details of the applied feedback and ISM model.
For the model and the resolution applied in this work, we use a value of $n_{\rm{th}}\sim 3 \;\rm{cm}^{-3}$, for
which we find the best results. As argued by \citet{eris}, a significantly lower value leads to less efficient
feedback and higher bulge fraction. However, a significantly higher threshold in our model leads to the formation
of bound stellar clumps, which can sink to the centre because of dynamical friction and thus also enhances the 
bulge fraction. Our value lies between the value of $1 \;\rm{cm}^{-3}$ applied for the Gasoline model for halo Aq-C-5
in CS12 and $5 \;\rm{cm}^{-3}$ applied by \citet{eris} in a higher resolution simulation.
We note that these values are still two orders of magnitude below the average density of molecular clouds
and at least four orders of magnitude below the density of molecular cores within which stars are observed
to form. Our resolution is however too coarse to model these high densities.

For particles with $n>n_{\rm{th}}$ and an overdensity $\rho/\bar{\rho}>2000$ 
(where $\bar{\rho}$ is the cosmic mean density) which lie in a convergent flow, a star formation rate of
\begin{equation}
{{\rm{d}\rho_{\star}} \over {\rm{d}t}} \, = \, \eta \, {{\rho_{\rm{gas}}}\over{t_{\rm{dyn}}}}
\end{equation}
is assumed. Here stellar and gas densities are represented by $\rho_{\star}$  and $\rho_{\rm{gas}}$ and $t_{\rm{dyn}}=
1/\sqrt{4 \pi G \rho_{\rm{gas}}}$ is the local dynamical time for the gas particle. We choose a star formation efficiency
$\eta=0.04$, in the range of values typically used (see e.g. various models in CS12). The typical
timescale for star formation is thus $t_{\rm{sf}}=1/\eta \; t_{\rm{dyn}}$. Star particles are created stochastically
with one gas particle being turned into one star particle of the same mass.

\subsection{Metal Production and Cooling}
\label{metals}

To account for metals, we explicitly trace the mass in the elements H, He, C, N, O, Ne, Mg, Si, S, Ca and Fe
for all gas and star particles. Our model includes chemical enrichment from SNII, SNIa and AGB stars.
Each star particle represents a stellar population characterized by a \citet{kroupa} IMF with lower
and upper mass limits of $0.1$ and $100 \; M_{\odot}$.

We assume that stars more massive than $8 \; M_{\odot}$ explode as SNII. All SNII are modeled in one event at an age $\tau_{\rm SNII}$
(see Section 2.4). For the calculation of the mass returned to the ISM in the various elements
we use the metal-dependent yields provided by \citet{chieffi}. The uncertainties in yields and thus in the predictions
of simulations are significant, as was for example discussed by \citet{wiersmaM}. It has been argued by \citet{portinari}
that adjustments by factors of 2 for certain elements can help improve the agreement with observations. Indeed, we find that
halving the iron yield following their findings leads to a qualitatively better agreement of metallicities as discussed
in the following sections, which is why we apply their suggested corrections. We note that apart from this, we have not
studied the variation of our results with different yield sets, IMFs etc. to optimize our results.

For the element production by SNIa, we assume the model W7 presented by \citet{iwamoto}. We
apply a delay time distribution, which declines with age $\tau$ of a stellar population as $\tau^{-1}$,
as proposed by \citet{maoz}. We also adopt their suggested normalization of 2 SNIa per 1000 $M_{\odot}$
of stars formed and that the first SNIa explode at $\tau = 50 \; \rm{Myr}$. The corresponding
masses of elements are returned to the ISM in $50\;\rm{Myr}$ time-steps in our model.

To account for the mass recycling in the winds of asymptotic giant branch stars, we use the metal-dependent
yields of \citet{karakas}. Together with the assumed IMF and using lifetimes dependent on stellar mass and metallicity, we can thus
calculate the mass of the considered elements released during the time intervals considered
for the SNIa enrichment.

Chemical elements are distributed to the gaseous neighbours of the star particles, where neighbours
are weighted according to their distance from the star particle using an SPH kernel. To account for 
our multiphase treatment of the gas component, the returned (metal) mass is split between 10 hot and 
10 cold neighbours. We give 50 per cent to the hot and cold phase each for all three different types
of metal production sites. We have tested making this fraction dependent on the age of the stellar
particle, but found that observed abundance ratios are best reproduced by our simple choice.
For this purpose the cold phase gas is defined by $T<8\times10^{4}\;\rm{K}$ and $n>4.\times10^{-5}\;\rm{cm}^{-3}$.
Note, that we have reduced this density limit by a factor of $\sim 100$ compared to \citet{cs06}
as we found that a higher value can have a destructive effect on extended, low-density gas disks
due to energetic feedback (see below).

The metallicities of the SPH particles are used to calculate the cooling rates of the gas.
We apply the rates presented by \citet{wiersmaC} for optically thin gas in ionization equilibrium. 
These rates are calculated on an element-by-element basis and take into account the effects
of photo-ionization from a uniform redshift-dependent ionizing background \citep{haardt}. The rates
thus depend on redshift, gas density, temperature and chemical composition.

\subsection{Metal Diffusion}
\label{metdif}

In many standard implementations of chemical enrichment in SPH (also true for \citealp{cs05}), the metallicity of a particle
can only change by enrichment. This can lead to situations, where gas particles with similar thermodynamic properties,
but very different metallicities live next to each other. This occurs e.g. when a galactic wind particle travels through unenriched IGM.
\citet{wiersmaM} suggested that smoothed metallicities should be used to get rid of these situations. Although
this improves the modeling by smoothing out differences, it also leads to fluctuating metallicities for individual particles
for the IGM situation discussed above.

In the ISM, once metals have been released from stars, turbulent motions of gas are responsible for their spreading.
Including a corresponding model for the turbulent diffusion of metals was already suggested by \citet{groom},
but until recently, most galaxy formation SPH codes did not consider this process. Recent implementations were presented in
\citet{martinez}, \citet{greif} and \citet{shen}.

The diffusion equation for a metal concentration $c$ (metal mass per total mass) of a fluid element with density $\rho$
and a diffusion coefficient $D$ is
\begin{equation}
{{\rm{d}\it{c}}\over{\rm{d}\it{t}}}\;=\;{{1}\over{\rho}}\nabla \cdot (D\nabla c),
\end{equation}
where d/d$t$ is the Lagrangian derivative. \citet{cleary} gave the SPH formulation of the diffusion equation as
\begin{equation}
{{\rm{d}\it{c_i}}\over{\rm{d}\it{t}}}\;=\;\sum\limits_{j} K_{ij} \left( c_i - c_j \right),
\end{equation}
where
\begin{equation}
K_{ij}\;=\; {{m_j}\over{\rho_i\rho_j}} {{4D_iD_j}\over{(D_i+D_j)}} {{{\bf r}_{ij}\cdot \nabla_i W_{ij}}\over{r_{ij}^2}}.
\end{equation}
Here quantities with subscripts {i} and {j} correspond to neighbour particles, $m$ is the particle mass,
$W_{ij}$ is the SPH kernel and ${\bf r}_{ij}$ is the separation vector with absolute $r_{ij}$.

\citet{greif} argued for the use of an integrated equation assuming that the change in $c$ is small over $\Delta t$:
\begin{equation}
c_i(t_0+\Delta t)\;=\;c_i(t_o)e^{A\Delta t} + {{B}\over{A}}(1-e^{A\Delta t})
\end{equation}
with
\begin{equation}
A=\sum\limits_{j} K_{ij} \,\,\,\,\,\,\, {\rm and} \,\,\,\,\,\,\,\, B=\sum\limits_{j}K_{ij}c_{j}.
\end{equation}
As we want to conserve the total metal mass, we modify the equation for a pairwise exchange of metals.
For the metal mass $\mu_{i}=c_i m_i$ we get
\begin{equation}
\Delta \mu_{i} = \sum\limits_{j}{\mu_{ij}}=\sum\limits_{j} \left[
{{1}\over{2}}m_i\left(1-e^{A\Delta t}\right){{1}\over{A}}K_{ij}\left(c_j-c_i)\right)\right],
\end{equation}
where the factor 1/2 was included to account for the fact that most pairs of neighbours are considered twice
and $\mu_{ij}$ is correspondingly subtracted from particle $j$. To avoid dependence on the ordering of particles
all changes $\Delta \mu_{ij}$ are calculated for all pairs of neighbours before the metal masses of all particles are updated.
This procedure is applied at every time-step for all active particles using the standard SPH neighbour searches. 
A neighbour particle $j$ can be inactive, so that the corresponding pair of particles only appears once.
Should the particles still be neighbours at the next active time-step of $j$, the larger $\Delta t$ compensates for that.
Clearly this formalism includes a number of approximations, but applied to tests as discussed in \citet{greif},
we find a similar accuracy.

This leaves the determination of the diffusion coefficient $D$. \citet{greif} argued for $D_i=2\rho_i\sigma_i h_i$, where
$\sigma$ is the velocity dispersion of gas particles within its smoothing kernel characterized by the smoothing length $h_i$.
\citet{shen} argued that $D_i=0.05 \rho_i \left| S_{kl} \right| h_i^2$ based on the trace-free tensor $S_{kl}$ (for details
see their paper) is a better choice as it yields no diffusion for purely rotating or compressive flows.
We have tested both ideas and found that for a fixed test setup the main difference is the strength of the diffusion coefficient
with \citet{greif} predicting values higher by a factor $\sim 20$. For cosmological simulations, we find that the Shen et al. configuration
yields better results when comparing to observations. 
As was noticed by \citet{shen}, diffusion leads to outflowing particles losing metals to the circumgalactic
medium and subsequently also to higher gaseous and stellar metallicities in the galaxy. For the Greif et al. value for $D$
this effect is much stronger than for the Shen et al. value and makes galaxies lie above the mass-metallicity relation.
However when we use $D_i=0.1\rho_i\sigma_i h_i$ this criterion loses significance.
For consistency, we use $D$ as suggested by Shen et al. in the simulations presented in this paper.

\subsection{Thermal and Kinetic feedback}

As in \citet{cs06} we assume that each SN ejects an energy of $\sim 10^{51} \rm{erg}$ into the surrounding ISM.
As for the metals we split this energy in halves and give those to the 10 nearest hot and cold gas neighbour particles
as defined above. However, we now split the energy between a kinetic and a thermal part (see also \citealp{agertz13}). 

To determine the kinetic part, we consider the conservation of momentum $\Delta p= \Delta m\; v_{\rm out}$ contained in the initial SN ejecta
and assume that this is characterized by a typical outflow velocity $v_{\rm out}$. We use $v_{\rm out}=3000 \; {\rm kms}^{-1}$,
a typical velocity of outflowing material in SN in the Sedov expansion phase.
Note that the kinetic energy carried by $\Delta m = 10 \;M_{\odot}$ at this velocity 
is $0.9\times10^{51} \rm{erg}$, the average over all SNII according to our choice of IMF and SNII mass interval is $\sim1.5\times10^{51} \rm{erg}$.
With this assumption, the momentum transferred from a star particle to one of the 20 gas neighbour particles $i$ receiving feedback
is determined by $\Delta m_{i}$, the mass transferred to particle $i$, which we know from the considerations of Section 2.2
for each particle $i$ at a given feedback time-step. The momentum transferred to particle $i$ is simply $\Delta p_{i} = \Delta m_{i}\; v_{\rm out}$.
The direction of the momentum change vector is modeled as pointing radially away from the star particle towards
gas particle $i$. Our choices of parameters lead to a typical change in radial velocity component
of a gas particle $i$ receiving feedback of $\Delta v_{i} =20-30 \;\rm{kms^{-1}}$. As $\Delta m$ is not spread equally among particles, in extreme cases
$\Delta v_{i} \sim 100 \;\rm{kms^{-1}}$ is possible.

We also know how many SN events are represented and thus the total energy that is available to be released, $\Delta E_{\rm tot}$,
under the assumption that per SN an average of $1.0\times10^{51} \rm{erg}$ is transferred to the ISM.
The transfer of momentum leads to a change in kinetic energy of the gas particle $\Delta E_{\rm kin}$. 
The remaining energy is considered to be thermalized, $\Delta E_{\rm{therm}}= \Delta E_{\rm tot}-\Delta E_{\rm kin}$. For this energy we follow the
ideas of \citet{cs06}. The fraction transferred to a hot particle is instantly added to its thermal energy. Instead, for cold particle
we accumulate the energy from SNe events in a reservoir. Only when the accumulated energy is high enough, so that it can become a hot
particle, the energy is released ('promotion'). To define 'hot' in this context, we search for neighbour particles with entropies high enough to be 
decoupled from the cold particle in our multiphase scheme and calculate their mean entropy $A_{\rm hot}$. 
If a particle has less than 5 such neighbours within 10 smoothing lengths,
we set $A_{\rm hot}=A_{\rm{th}}$, where $A_{\rm{th}}$ corresponds to $T=1.6\times10^{5}\;\rm{K}$ and $n=2\times10^{-3}{\rm cm}^{-3}$
('seeding', see \citealp{sawala}). Moreover, $A_{\rm hot}=A_{\rm{th}}$ is also assumed if $A_{\rm hot}<A_{\rm{th}}$, so that
$A_{\rm{th}}$ acts as a minimum promotion entropy.

As $\Delta E_{\rm{therm}}> \Delta E_{\rm kin}$ for the considered situation, the thermal feedback is only mildly weakened.
The kinetic feedback however helps breaking up dense clumps of gas and thus lowers the cooling rates and
the star formation efficiency in a star-forming disk. Gas fractions increase and so does the efficiency of the thermal feedback.
Momentum is also transferred, when mass is released from AGB stars, however masses and velocities are much smaller than in
SN explosions, which is why this effect is negligible.

For the age of star particles at which we input SN energy we choose $\tau_{\rm SNII} = 3\; \rm{Myr}$.
Clearly this choice is not unproblematic, as most SNII explode after that, but the energy release from 
SNII peaks at this time and the most massive stars are supposed to have the strongest effect on the ISM. When
not using feedback from massive stars before their explosion as SN as described in the following section, we find
less realistic star formation histories for a later choice of $\tau_{\rm SNII}$. When adding an additional source of feedback
this effect becomes less significant. Star formation rates for individual haloes can go up or down by up to a factor of 2
at a given epoch for a given halo when $\tau_{\rm SNII}$ is increased, but cumulative effects are weaker.
From these test, we conclude that the choice of $\tau_{\rm SNII}$ does not significantly impact the results of our paper.

\subsection{Radiation Pressure}
\label{RP}

\citet{hopkins} used high-resolution simulations to test the idea that the inclusion
of feedback from radiation pressure of young massive stars
has a comparable or possibly even stronger effect on the ISM than SN feedback (for the underlying ideas see e.g. \citealp{murray}).
They developed a model for high resolution 
simulations, in which they identify star forming regions and can thus use their properties to calculate the effect 
of radiation pressure as a function of these local properties. \citet{stinson13} have introduced a model, where
instead of a momentum transfer, a thermal energy transfer is modeled, assuming that 10 per cent of the radiated energy from
massive young stars is thermalized in the surrounding ISM. \citet{agertz13} presented a model for radiation pressure, which
assumes that each star particle represents a sample of star forming clusters. The effect is however then dependent
on their choice of the maximum cluster mass.

We parametrize the rate of momentum deposition to the gas as
\begin{equation}
\dot{p}_{\rm rp}=(1+\tau_{\rm IR}){{L(t)}\over{c}},
\end{equation}
where $\tau_{\rm IR}$ is the infrared optical depth and $L(t)$ is the UV-luminosity of the stellar population and
we have set additional efficiency parameters to 1 (see \citealp{agertz13}). 
This equation states that all UV photons are scattered or absorbed by dust,
which subsequently re-radiates the energy in the infra-red.
The IR photons are then scattered multiple times before leaving the star forming cloud.
We construct $L(t)$ by using the stellar evolution models for massive stars
by \citet{ekstrom} and assuming a \citet{kroupa} IMF.

\begin{table}
  \centering
    \begin{tabular}{@{}rcccc@{}}
      \hline
      Halo & $M_{\rm{vir}}$ & $m_{\rm{dm}}$ & $m_{\rm{gas}}$ & Origin \\
      & [$10^{10} M_{\odot}$] & [$10^{6} M_{\odot}$] &[$10^{5} M_{\odot}$] &\\
      \hline
      \hline
      6782-4x&  17.03 & 3.62 & 7.37 & LO\\
      4323-4x&  29.50 & 3.62 & 7.37 & LO\\
      4349-4x&  30.28 & 3.62 & 7.37 & LO\\
      2283-4x&  49.65 & 3.62 & 7.37 & LO\\
      Aq-B-5 &  70.35 & 1.50 & 2.87 & CS\\
      1646-4x&  81.61 & 3.62 & 7.37 & LO\\
      1192-4x& 100.03 & 3.62 & 7.37 & LO\\
      1196-4x& 113.81 & 3.62 & 7.37 & LO\\
      0977-4x& 129.56 & 3.62 & 7.37 & LO\\
      Aq-D-5 & 150.43 & 2.31 & 4.40 & CS\\
      Aq-C-5 & 151.28 & 2.16 & 4.11 & CS\\
      Aq-A-5 & 164.49 & 2.64 & 5.03 & CS\\
      0959-4x& 164.54 & 3.62 & 7.37 & LO\\
      0858-4x& 182.44 & 3.62 & 7.37 & LO\\
      0664-4x& 213.74 & 3.62 & 7.37 & LO\\
      0616-4x& 235.77 & 3.62 & 7.37 & LO\\
      \hline
      \hline
    \end{tabular}
    \caption{Overview over the haloes studied in this paper,
             including the name of the halo, its virial mass
             $M_{\rm{vir}}$, the dark matter particle mass
             $m_{\rm{dm}}$ and the initial gas particle mass
             $m_{\rm{gas}}$. Under 'Origin' we distinguish
             between haloes from 'CS' (CS09) and 'LO'
             \citep{oser}.
             }
    \label{overview}
\end{table}

\begin{figure}
\centering
\hspace{-0.65cm}\includegraphics[width=9cm]{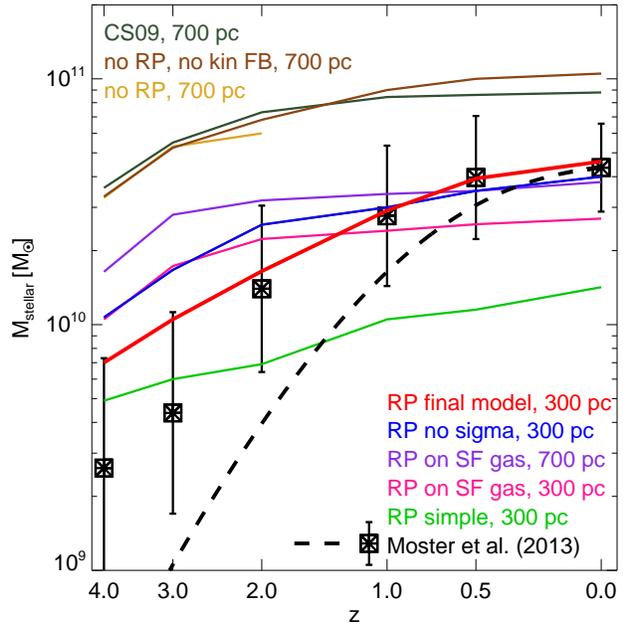}
\caption{The evolution of the stellar galaxy mass $M_{\rm stellar}$ vs. redshift $z$ in halo AqC for various 
code configurations (coloured lines). Overplotted are the predictions of MNW13 in two ways:
a) The predicted mass evolution of a typical halo with the $z=0$ mass of AqC (black dashed line) and b) the predicted
evolution taking into account the actual halo mass of AqC at six redshifts (points with 1 $\sigma$ error-bars).}
\label{aqc}
\end{figure}

More problematic is the estimation of $\tau_{\rm IR}$. If we follow \citet{hopkins}, $\tau_{\rm IR}=\Sigma_{\rm clump}\kappa_{\rm IR}$
with $\kappa_{\rm IR}\sim5\;\rm{cm^2g^{-1}}$, we effectively have to know the surface density of the star forming gas clump.
Due to our coarse resolution, we do not resolve such clumps. We therefore use the following model:
\begin{equation}
\tau_{\rm IR} = \tau_0 f(\rho, Z)g(\sigma).
\end{equation}
Each gas particle is characterized by its density and metallicity through $f(\rho, Z)$, whereas the environment of the star particle
is characterized by $g(\sigma)$.  $f(\rho, Z)$ basically models the dependence on the dust surface density represented by a particle:
\begin{equation}
f(\rho, Z) = max \left[ 1,\left({{\rho h}\over{\rho_{\rm th} h_{\rm th}}}\right)\right]\left({{Z}\over{Z_{\odot}}}\right)
\end{equation}
Here we use $\rho h$ with the smoothing length $h$ as a measure for the surface density represented by a particle and limit the effect
of this factor at the star formation density threshold. We also assume a linear dependence of dust on metallicity $Z$.
If we set $g=1$, we gain insight into how our model works. If we choose $\tau_0=25$, which is a value in the range of values
10-100 found by \citet{hopkins} in their models of star forming high-$z$ disks, we find that indeed radiation pressure significantly
reduces high-$z$ star formation rates and brings them into better agreement with abundance matching results, similar to the findings
of \citet{stinson13}. However, the low-$z$ star formation efficiencies become too low and the final galaxy masses are too small
indicating that radiation pressure is too strong at later times
(see Figure \ref{aqc} and discussion in Section \ref{aqcsec}).

However low-z disks show significantly lower gas velocity dispersions than gas-rich, turbulent high-$z$ disks \citep{genzel}.
For a typical Milky-Way type halo, we find the average gas velocity dispersion $\sigma$ is $\sim 40 \; {\rm kms}^{-1}$ at $z\sim 2$, which is
in agreement with observed galaxies at that time \citep{fs09},
and at low redshift $\sigma \sim 15-20 \; {\rm kms}^{-1}$, which is too high, but explained
by our coarse resolution. Star forming regions, and thus the values of $\tau_{\rm IR}$, are thus much larger at high $z$.
If we assume that the Jeans mass is a valid estimator for clump mass, then the mass scales as $M_{\rm cl}\propto\sigma^3$
(we ignore the dependence on density as we are not modeling densities above the star formation threshold properly). 
As is pointed out in the review of observed star forming clouds in \citet{agertz13}, the final radius of clusters depends weakly on mass,
which leads to our crude estimation
\begin{equation}
g(\sigma)=\left( {{\sigma}\over{\sigma_0}}\right)^3.
\end{equation}
To estimate $\sigma$ in the environment of the star particle, we determine the velocity dispersion $\sigma_{\rm kernel}$ 
of gas particles within the smoothing kernel for each gas particle. Then we average over the values $\sigma_{\rm kernel}$ 
of all the neighbour gas particles to a young star particle to avoid strong fluctuations.

We acknowledge that this line of estimation is not particularly stringent, but it qualitatively assures that the effect of
radiation pressure is stronger in galaxies with higher gas velocity dispersions and thus larger star forming regions.

In our simulations, we model the momentum change $\Delta p = \dot{p}_{\rm rp} \Delta t$ where $\Delta t$ is the time-step of the
star particle, as a continuous force acting on the 10 nearest cold neighbour particles during the first 30 Myr in the life of a star 
particle. We use $\tau_0=25$ for $\sigma_0=40 \; {\rm kms}^{-1}$, so that the effect at high-$z$ reduces star formation rates
and is weak at low-$z$. We also limit $g(\sigma)$ at a value of 4 to avoid overly strong forces. 
By construction, we thus find values of $\tau_{\rm IR}\sim 20$ at $z\sim 2$ with extreme values going up to $100$, whereas
we find significantly lower values of the order $1-5$ at low redshift. 
We discuss the choice of parameters in Section \ref{aqcsec}.

\section{The sample}
\label{sample}

As initial conditions for our cosmological zoom-in simulations we use haloes from the Aquarius Project
(\citealp{aquarius}, CS09), one of which was also used for the \textit{Aquila Project} (CS12).
In addition, we use a selection of haloes from \citet{oser}. 

The Aquarius haloes are a suite of high resolution zoom-in resimulations of regions chosen from the Millenium II simulation \citep{mii} which follows
a cosmological box of a side-length $137\; \rm {Mpc}$. They were simulated from $z=127$ assuming a 
$\Lambda$CDM universe with the following parameters: $\Omega_{\Lambda}=0.75$, $\Omega_m=0.25$,$\Omega_b=0.040$,
$\sigma_8=0.9$ and $H_0=73\;\rm{kms}^{-1}\rm{Mpc}^{-1}$. The haloes were selected to have a similar mass
to that inferred for the Milky-Way dark halo and to have no neighbour exceeding half of their mass within 
$1.4 \; \rm{Mpc}$. For details we refer to \citet{aquarius} and CS09.
Because of the second criterion, the haloes have a relatively quiet low-$z$ merger history.
They are thus expected to host galaxies with high disk fractions at $z=0$. Disks were indeed found to form in these haloes by CS09 and
CS12, but with low disk-to-bulge ratios and several other properties that do not compare well to observations.

The haloes from \citet{oser} were selected from a simulation of a cosmological box with a side-length of 
$100\; \rm {Mpc}$, and were simulated from $z=43$ assuming a $\Lambda$CDM universe with the following parameters:
$\Omega_{\Lambda}=0.74$, $\Omega_m=0.26$, $\Omega_b=0.044$, $\sigma_8=0.77$ and $H_0=72\;\rm{kms}^{-1}\rm{Mpc}^{-1}$.
Unlike the Aquarius haloes, these haloes were not chosen specifically as candidates for disk galaxies and the sample
thus also contains $z\sim0$ mergers. For details we refer to \citet{oser}, who resimulated lower resolution
versions of these haloes.

The haloes we selected from these two samples are listed in Table \ref{overview}.
Our combined sample comprises 16 haloes with $z=0$ re-simulated virial masses $M_{\rm{vir}}$ ranging from $1.7\times
10^{11} M_{\odot}$ to $2.4\times 10^{12} \; M_{\odot}$. The resolutions of the simulations are characterized by the
initial gas particle masses, which lie between $2.87$ and $7.37\times 10^{5} \; M_{\odot}$ and
dark matter particle masses between $1.50$ and $7.37\times 10^{6} \; M_{\odot}$.
For our re-simulations we applied comoving gravitational softening lengths of $h_{\rm{bar}}\sim 300 \; \rm{pc}$ for gas
and stars, and of $h_{\rm{dm}} \sim 650 \; \rm{pc}$ for dark matter. We thus use shorter
softening lengths than in CS09.

\section{Star formation histories}

We begin to analyze the outcome of our simulations by considering the star formation history (SFH) of the model galaxies.
As has been shown by \citet{moster} (MNW13 hereafter), the majority of simulations predicts stellar galaxy masses
that are significantly higher than abundance matching results suggest at all redshifts $z=0-4$ with
overproduction being stronger at high redshift (for exceptions see e.g. \citealp{stinson13, kannan}).
As real disk galaxies form most of their stars at low-$z$, when destructive mergers are rare, it is crucial 
for simulations to produce reasonable SFHs in order to get galaxy structural properties right.

\begin{figure}
\centering
\hspace{-0.7cm}\includegraphics[width=9cm]{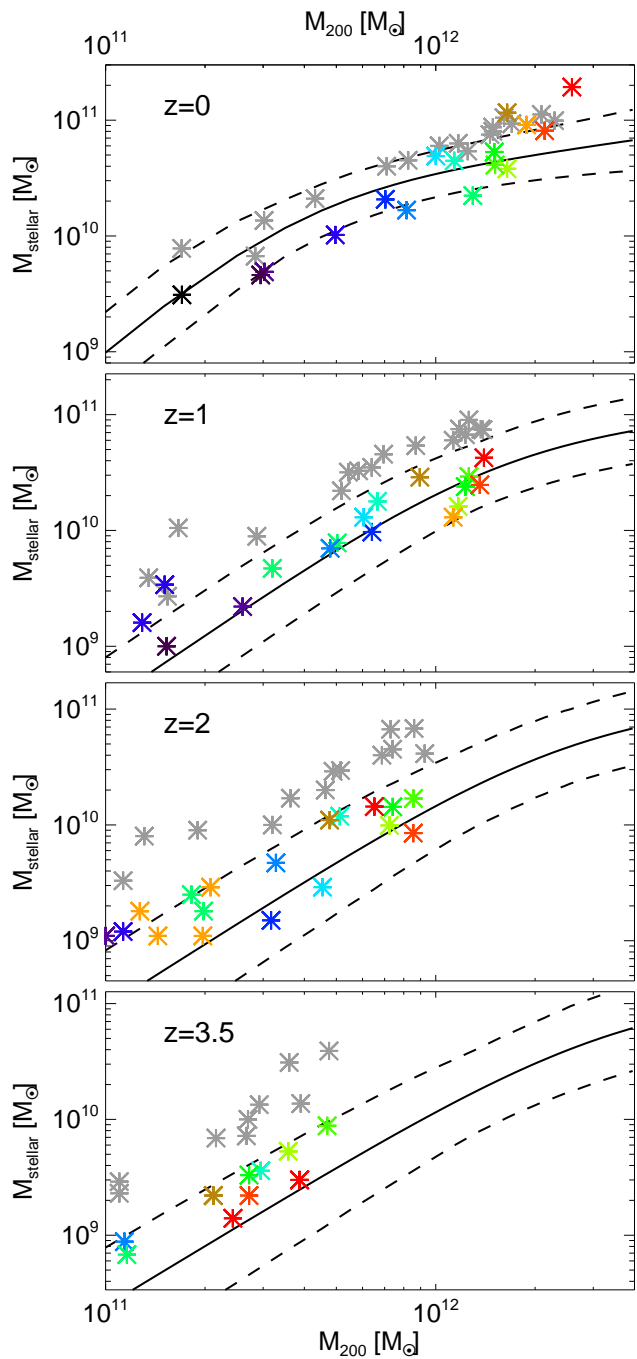}
\caption{Stellar galactic mass $M_{\rm stellar}$ plotted against halo virial mass $M_{200}$ at redshifts $z=0,1,2$ and $3.5$.
The coloured points are the models discussed in this paper. Each colour represents a simulation. A colour can appear more than
once at high-$z$ representing progenitor haloes. This colour coding is used for figures throughout the paper.
The gray points are models of the same set of haloes run with the code used in CS09. 
The black lines are the abundance matching results of
MNW13 with $1\sigma$ regions indicated by the dashed lines.}
\label{moster}
\end{figure}

\subsection{The effect of changing feedback models on the SFH of AqC}
\label{aqcsec}

Halo AqC has been studied in CS12 with a variety of different simulation codes. Although some of these codes
produce realistic $z=0$ masses, none produced a good match to abundance matching results for the SFH at all $z=0-4$.
Compared to haloes of similar $z=0$ mass, AqC is among the ones which assemble their dark mass
earliest (see Figure A1 in \citealp{cs11}). When developing our feedback models, we found that AqC is particularly sensitive to the choice
of model details, which we demonstrate in this section. We caution, that the effects we discuss here
can vary from halo to halo and not all conclusions drawn from AqC are true for all haloes. 
Moreover, abundance matching gives statistical properties for a galaxy sample and AqC could well be an outlier.
As we will show below, our calibration method is justified by the fact that
the feedback model that works best for AqC produces a sample of galaxies with reasonable properties.

\begin{figure*}
\centering
\includegraphics[width=18cm]{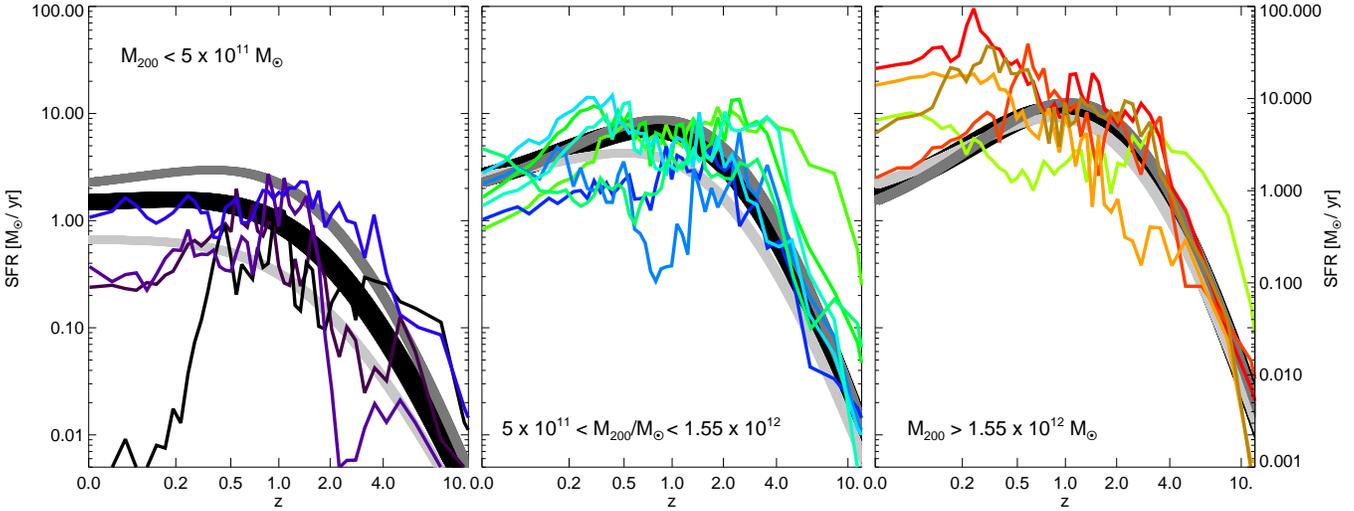}
\caption{Star formation rates $SFR$ as a function of redshift $z$ for all our models. 
The haloes are split into $M_{200} < 5\times 10^{11} M_{\odot}$ (left),
$5\times 10^{11} M_{\odot} < M_{200} < 1.55 \times 10^{12} M_{\odot}$ (middle) and $M_{200} > 1.55\times 10^{12} M_{\odot}$ (right).
Overplotted are predicted SFRs for typical haloes given by MNW13 for the mean, lowest and highest halo mass per panel.}
\label{sfr}
\end{figure*}

In Figure \ref{aqc} we plot the evolution of the stellar mass $M_{\rm stellar}$ from $z=4-0$ for various runs of the same initial conditions.
We compare it to the evolution predicted by the abundance matching results of MNW13, who provide two different
ways of comparison. On the one hand, fitting formulae for typical SF and accretion histories are presented as a function
of $z=0$ halo mass $M_{200}$. We represent these predictions by the dashed line in Figure \ref{aqc}.
On the other hand, we know the halo mass at each redshift $z$ and can thus make use of the fitting formulae for $M_{\rm stellar}/M_{200}(z)$.
We represent these by six data-points with corresponding error-bars from $z=4$ to $0$.
This method yields significantly higher predictions for $M_{\rm stellar}$ at high $z$, which reflects the
early assembly of halo AqC. Clearly the latter predictions should be used as a guideline for model calibration.

The original model by CS09 produces a stellar mass at $z=4$ that is about an order of magnitude too high
compared to the MNW13 value. It remains more than $1\sigma$ above predictions at all redshifts.
If we add our updated metal production and metal cooling rates (\textit{no RP, no kin FB}) the changes are
negligible. Changing to our new thermal and kinetic feedback (\textit{no RP}) also does not improve
the SFH at high redshifts (the run was stopped at $z=2$). 
We note that the change from the CS09 thermal feedback to our new thermal and kinetic feedback can produce
significant changes in star formation histories in other haloes.

\citet{stinson13} argued that adding feedback from young stars before their explosion as SNe would significantly shift
SF to later times. \citet{kannan} tested this idea on a simulation of a cosmological volume and found good agreement with the
results of MNW13 at high $z$. We tested various models of feedback from radiation pressure (RP)
First we consider a simple model for RP (\textit{simple RP}) where all affected gas particles are treated equally independent of density, metallicity
and gas velocity dispersion (assuming an effective optical depth $\tau_0=25$, see Equation 10). 
For this model we achieve a reduction of the early SFR so that the $z=4$ stellar mass
matches predictions. However, the feedback is too strong later on and SFRs are drastically reduced so that the $z=0$ stellar mass is more than 
$2\sigma$ low. When modifying our model so that the force is only acting on star-forming particles with $\rho > \rho_{\rm th}$
(\textit{RP on SF gas}), the strength of feedback is reduced especially at high $z$, so that the SFR is high at $z>4$, but the agreement at
$z=0$ is significantly better, however with low-$z$ SFRs being still too low. A model which includes a metallicity and density dependence
to account for the dust surface density of each gas particle as described in Section \ref{RP} (\textit{RP no sigma}) lowers
the effect of RP at low $z$ when metallicities are higher but the densities of the affected gas particles are on average lower, 
as the systems have lower gas fractions. Our full model including the dependence on the gas velocity dispersion $\sigma$ (\textit{RP final model})
models the dependence of RP on the size of star forming regions. As $\sigma$ is higher in high-$z$ galaxies, the effect of RP is
strengthened at high and weakened at low $z$ leading to a significant change in the shape of the $M_{\star}-z$ curve and bringing 
it within $1\sigma$ of the predictions of MNW13 at all $z<4$.

\citet{krumholz} have argued that the efficiency of the coupling between the radiation field and the dusty gas is significantly lower
than assumed in the approach of \citet{hopkins}, which is the basis of our model here. This would imply that radiation pressure is
less efficient in reducing the problem of overly high SFRs in simulated galaxies at high $z$. We thus caution, that the uncertainties
in the modeled processes remain high and we take the agreement of simulated SFH with the results of MNW13 as a justification for the
model we apply.

We also use Figure \ref{aqc} to show a dependence of our feedback on the gravitational softening length $\epsilon$. Going from 700 to 300 pc 
baryonic comoving softening length significantly reduces the high-$z$ SFRs, as shown by two versions of our \textit{RP on SF gas} model.
This seems to originate in a higher feedback efficiency due to denser structures in the very early stages of galaxy formation.
Because of this effect, we apply $\epsilon_{\rm baryons}=300\;\rm pc$ for our simulations, a significantly lower value than in CS09.
We note that reducing the softening length has no significant effect on the high-$z$ SFR in the absence of a model for radiation pressure.

We conclude that the inclusion of a parametrized model for feedback from 
young stars in the from of radiation pressure is indeed an efficient way to bring star
formation histories into better agreement with results from abundance matching techniques. We caution that details of the modeling
have a significant effect on the outcome and that these effects vary from halo to halo. It is crucial to test the model on a set of haloes,
as we describe in the next subsection.

\subsection{Applying the model to all haloes}
\label{rgal}

We now apply our updated galaxy formation code to the full sample of haloes described in Section \ref{sample} and examine if the resulting SFHs
agree with observations of the real galaxy population.

We start by discussing stellar masses $M_{\rm stellar}$ as a function of their halo masses $M_{200}$
at redshifts $z=3.5,2,1$ and $0$ in Figure \ref{moster}. 
$M_{\rm stellar}$ is defined as the mass of stars within a radius $r_{\rm gal}$ defined by visually analyzing the spherical
stellar mass profile $M_{\rm stellar}(<r)$ and determining where its radial growth saturates.
The coloured points in this Figure are for the models with our updated code, whereas the gray points are for a sample of simulations
run on the same set of initial conditions with the code version and model parameters applied in CS09.
Note that the colour coding according to halo mass that we apply here is used for all appropriate figures throughout the paper.
For the haloes 2283 and 0977, which are undergoing major mergers at $z=0$, we use the combined mass of the galaxies about to merge,
as their dark haloes have already merged.
At $z=0$, the population as a whole agrees nicely with the halo occupation modeling by MNW13. However, there are trends in the sample that disagree.
The four most massive haloes are $1-2 \sigma$ high, whereas the lower mass galaxies show a tendency towards overly low mass.

\begin{figure}
\centering
\hspace{-0.4cm}\includegraphics[width=9cm]{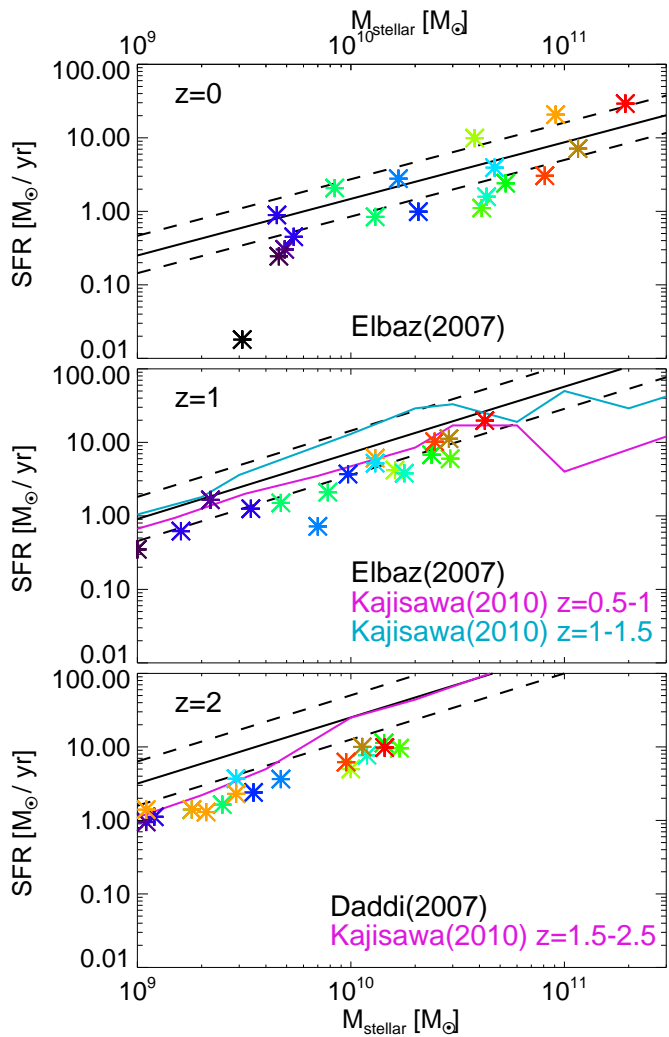}
\caption{Star formation rates $SFR$ at $z=0$, $z=1$ and $z=2$ (averaged over 1 Gyr each) vs. stellar galactic mass $M_{\rm stellar}$
compared to the observed relations found by \citet{elbaz} ($z=0,1$, black), \citet{daddi} ($z=2$, black) with $1\sigma$ errors
represented by dashed lines and to median relations found by \citet{kajisawa} (z=1,2, magenta and cyan).}
\label{sfrms}
\end{figure}

The sample simulated with the CS09 code shows $z=0$ stellar masses that
are generally too high, but in reasonable agreement with MNW13 at lower halo mass.
The strong discrepancy between the old code version and abundance matching results 
becomes better visible at higher redshifts, as in Figure \ref{aqc}.
The old code version yields stellar masses that for all haloes are more than $1\sigma$ high at $z\ge1$, as was also shown
for a large number of other codes in MNW13.

For the models simulated with our new code version the situation at $z\ge1$ is significantly different. Although
there is still a tendency towards too high mass at $z\ge2$, all values are within $1\sigma$ uncertainties of the
abundance matching results. At $z=1$, our models show perfect agreement with MNW13, indicating that the problematic
trends at $z=0$ originate only at low-$z$. We conclude that compared to the old code version and
to compilations of previous simulations (e.g \citealp{guo, sawala}, CS12, MNW13) our results look promising.

\begin{figure}
\centering
\hspace{-0.7cm}\includegraphics[width=9cm]{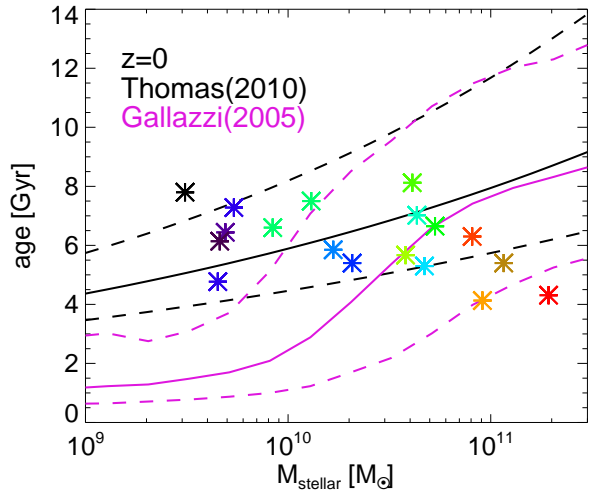}
\caption{Mean mass-weighted $z=0$ age of stars within 3 stellar half-mass radii $R_{50}$ vs. stellar galactic mass $M_{\rm stellar}$ 
compared to the observed relation found by \citet{gallazzi} for a magnitude-limited sample of galaxies (magenta lines).
We also overplot in black the relation for the luminosity-weighted age of elliptical galaxies by \citet{thomas}.
Dashed lines represent $1\sigma$ errors.}
\label{ages}
\end{figure}

We analyze the SFH problems further by separately plotting the SFHs of the most massive ($M_{200}>1.55\times10^{12}M_{\odot}$), least massive
($M_{200}<5.\times10^{11}M_{\odot}$) and remaining haloes in Figure \ref{sfr}.
We determine SFHs by following the in-situ growth of the main progenitor of the $z=0$ galaxy, so that accreted stars do not appear in the SFH.
To be consistent with Figure \ref{moster}, for the two $z=0$ major merger haloes, we add up SFHs of the two merging galaxies.
In each of the panels we overplot the predicted average SFHs for typical haloes of the lowest, mean and highest $z=0$ halo mass in the panel
as given in MNW13. These SFHs also do not include accreted stars.
As discussed above, individual halo assembly histories can differ significantly from the assumed average assembly histories.
The comparison can however reveal trends in the SFHs of our sample of simulated galaxies.

We start with the middle panel including AqC, which we have analyzed above. There is a trend for these haloes to overproduce stars at
$z>4$, with AqC showing the the highest SFRs, which are however mostly caused by its early assembly of mass.
Otherwise this trend agrees with the slightly high stellar masses at $z\ge2$.
As a population, these haloes show reasonable later agreement, leading to the agreement found in Figure \ref{moster} at $z\le1$.
Probably due to the high-$z$ overproduction, the predicted peak at $z=1-2$ is not well reproduced by the models. 
Moreover, there is a small tendency towards underproduction
at $z<1$. Generally, the conclusions for this sub-sample are the same as for AqC alone.

For the low mass haloes in the left panel, there is hardly any overproduction at high-$z$, which might be connected 
to the lack of resolution and the consequently unresolved star formation in small progenitor haloes. Tests
with higher resolution indicate that high-$z$ SFR increase with resolution for these haloes. The sub-sample
agrees with the predictions at $z\sim 0.5-5$, but late SFRs are low. This leads to the agreement with abundance matching at $z=1$, but
to the low $z=0$ masses seen in Figure \ref{moster}.

\begin{figure*}
\centering
\includegraphics[width=15.5cm]{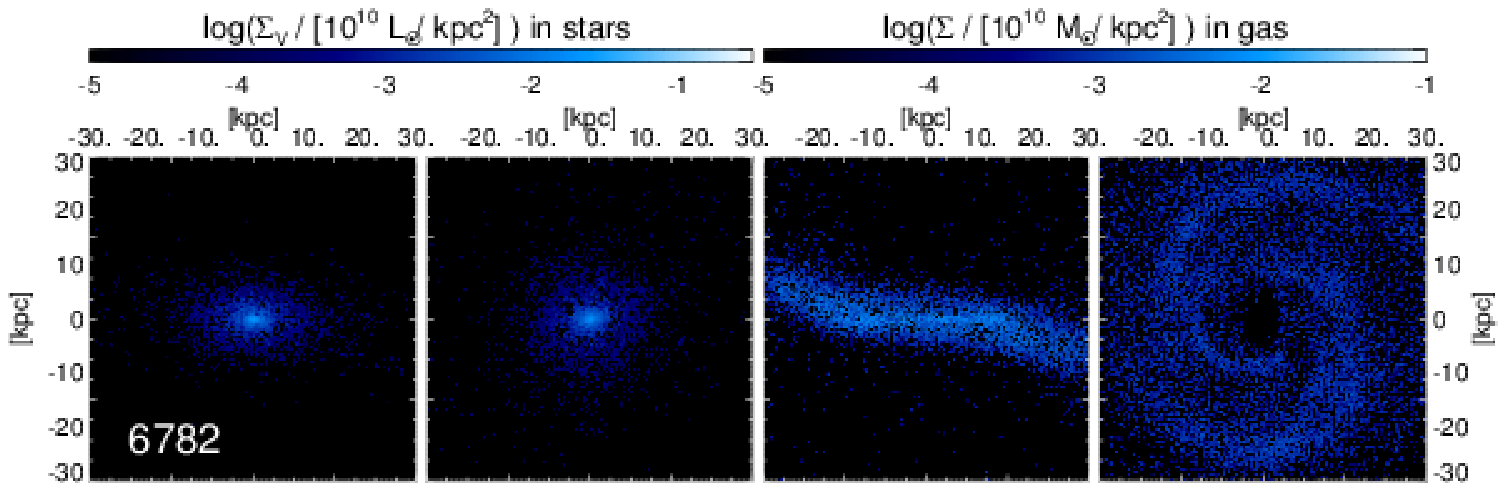}
\includegraphics[width=15.5cm]{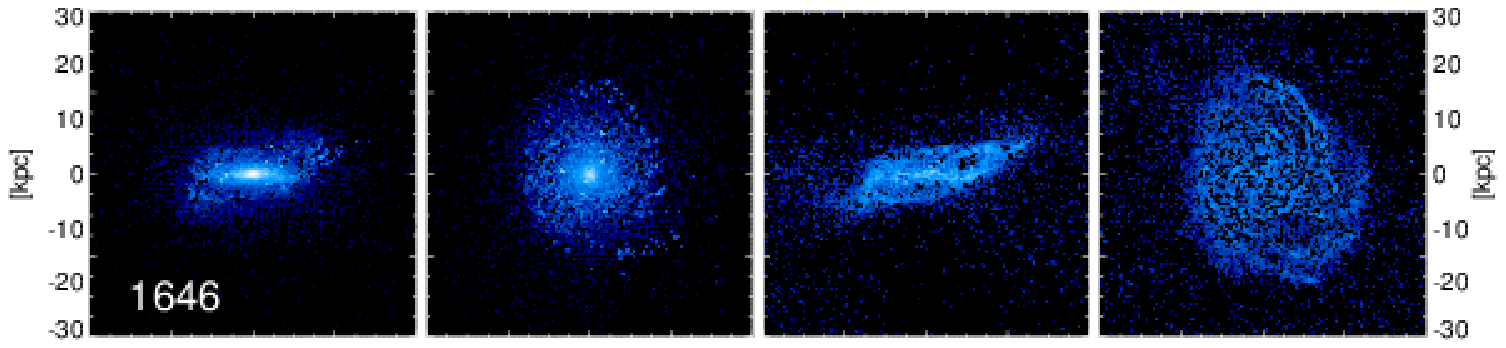}
\includegraphics[width=15.5cm]{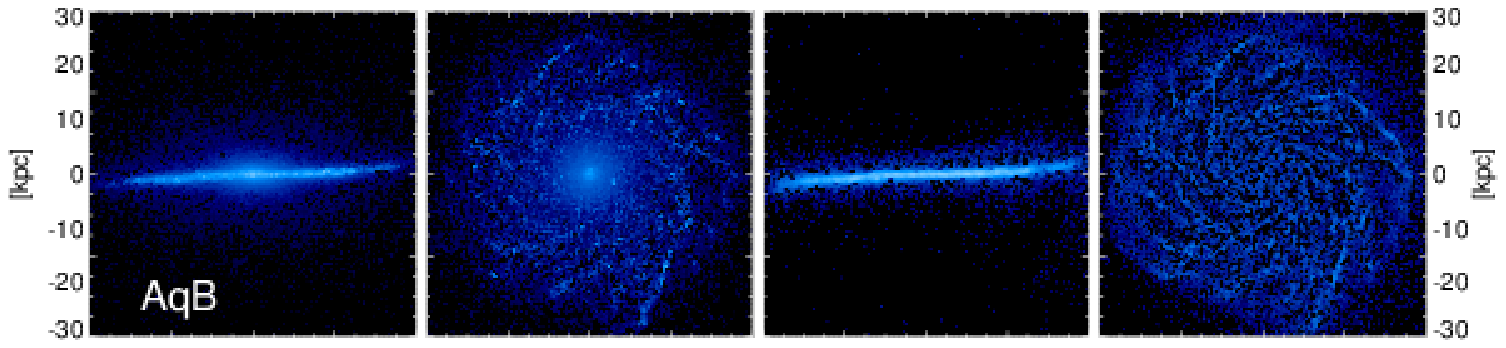}
\includegraphics[width=15.5cm]{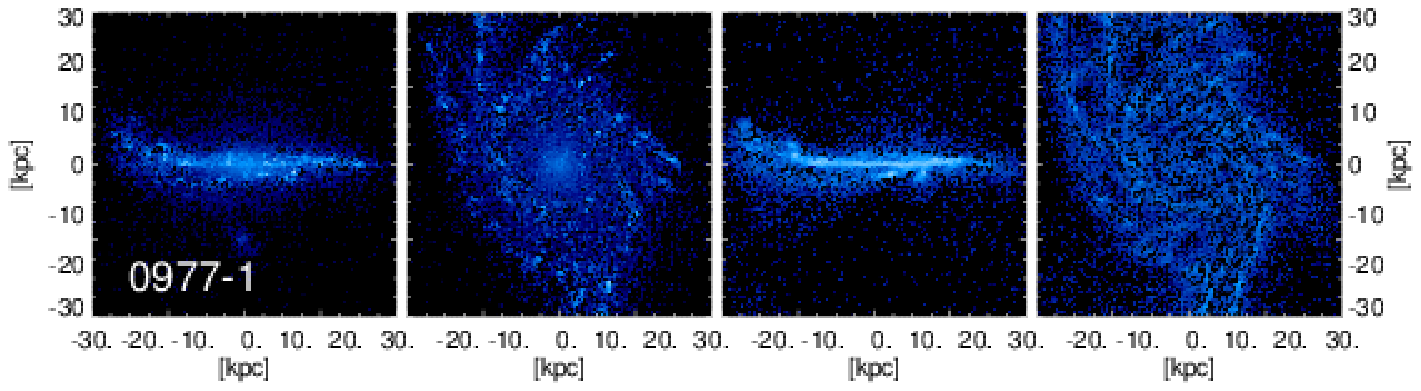}
\caption{Edge- and face-on images of the $z=0$ stellar $V$-band surface brightness $\Sigma_{V}$ (left) and of the gas surface densities $\Sigma$ 
 for our model galaxies in haloes
(top to bottom): 6782, 1646, AqB and one of the merging galaxies in 0977.}
\label{sd2d}
\end{figure*}

The high-mass haloes in the right panel show a significant overproduction of stars 
at high redshift only for halo AqA, which has a similar mass assembly history
to AqC. There is good agreement with predictions at $z\sim 0.8-4$. However, SFRs are significantly too 
high at later times and lead to high $z=0$ stellar masses. We find that at these times there is still efficient inflow of gas onto the galaxies,
but stellar feedback as applied in our models is too weak to efficiently remove gas from the disks as most of the ejected
gas returns in galactic fountains. AGN, which we do not model, are generally believed to stop hot halo gas from cooling
onto galaxies at these redshifts in massive haloes, though typically for higher masses than in our sample (e.g. \citealp{croton}).

\begin{figure*}
\centering
\includegraphics[width=15.5cm]{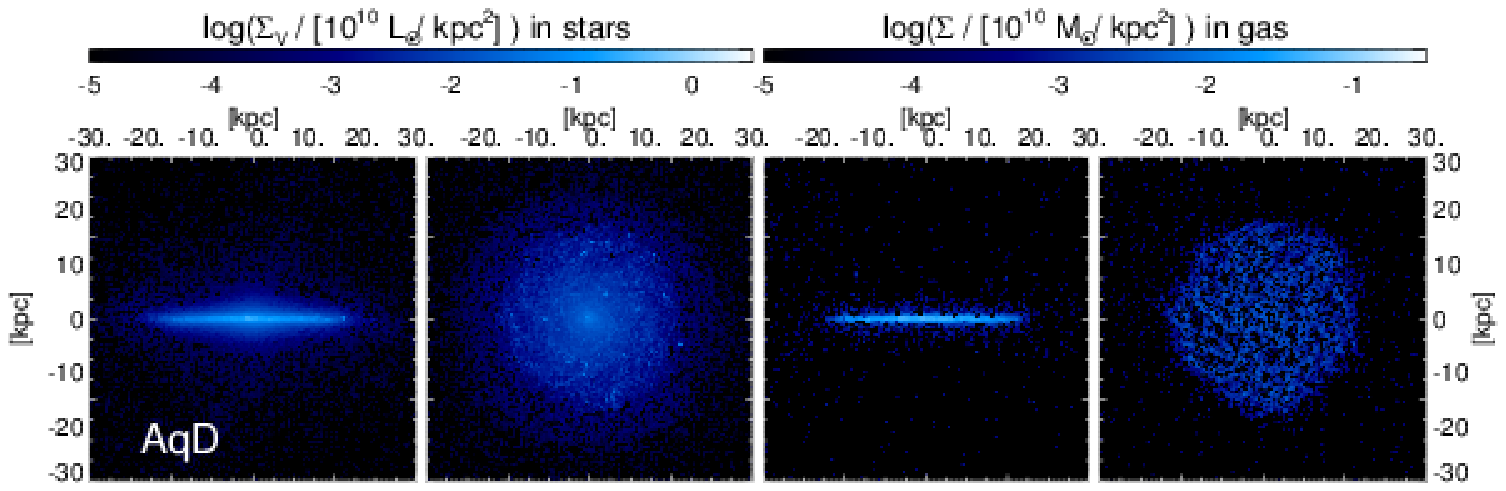}
\includegraphics[width=15.5cm]{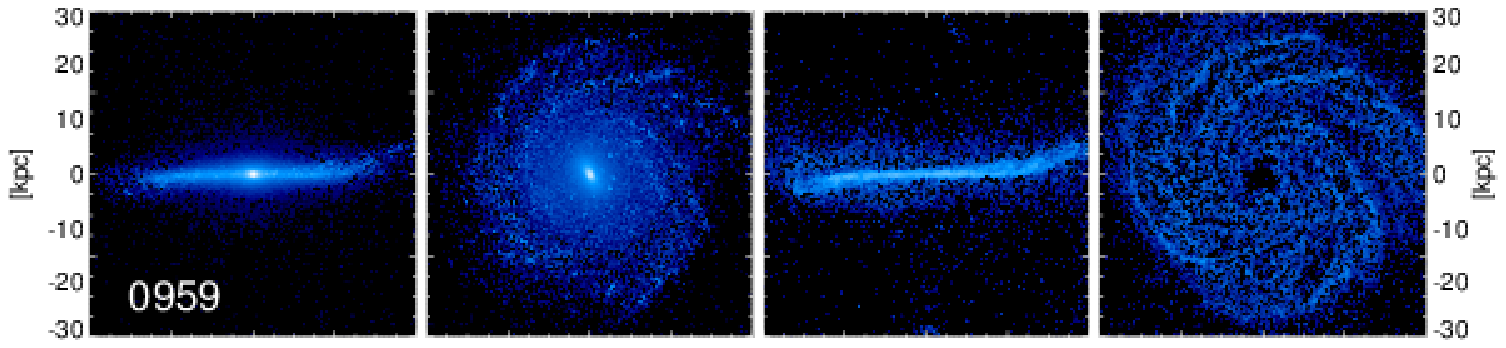}
\includegraphics[width=15.5cm]{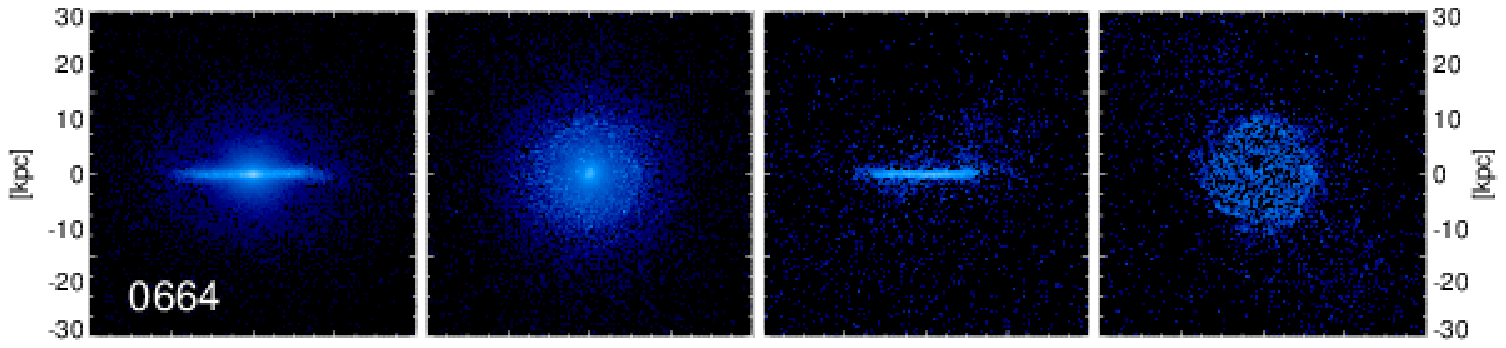}
\includegraphics[width=15.5cm]{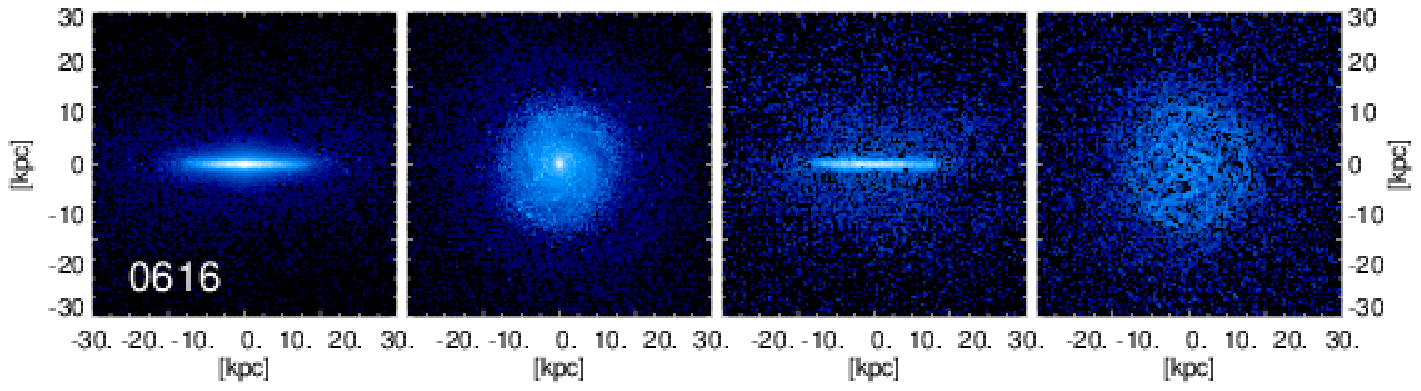}
\caption{As in Figure \ref{sd2d} for galaxies in haloes
(top to bottom): AqD, 0959, 0664 and 0616.}
\label{sd2d2}
\end{figure*}

To have a more direct comparison of our model SFRs to observation, we determine SFRs at $z=0,1$ and $2$ for Figure
\ref{sfrms} by averaging over 1 Gyr of star formation each. We compare the corresponding values to the relations found by \citet{elbaz}
and \citet{daddi} for observed SFRs vs. galaxy stellar masses.
For the major merger models 2283 and 0977 we here separately include 
the merging galaxies. Apart from our least massive halo 6782, in which there is hardly any star formation
after $z\sim 0.4$, our model sample agrees well with observation and all model galaxies lie within $2\sigma$ of the observed relation at $z=0$.
The high-mass haloes do show a tendency for high SFRs, as expected, and the rest of the models show a slightly lower average than 
observed, as was also concluded from Figure \ref{sfr}. It has to be noted, that the observations of \citet{elbaz} and \citet{daddi}
only considered blue, star-forming galaxies, whereas the results of MNW13 are an average over blue and red galaxies. 

At all three redshifts in consideration, the slope of the SFR vs. $M_{\rm stellar}$ relation
constituted by our models is in good agreement with observation.
At $z=1$, where the different observations agree well, SFRs are typically $\sim1\sigma$ low, 
which we can connect to the fact that our models do not reproduce the
peak in the SFHs at $z=1-2$ properly, but on average predict a flatter SFH. At $z=2$, our model SFRs are $\sim2\sigma$ low
compared to the results of \citet{daddi}. \citet{kajisawa} find lower SFRs at $M_{\rm stellar}<10^{10}M_{\odot}$,
which our lower mass galaxies agree with. The higher mass galaxies have significantly lower SFRs compared to these observations as well.
This seems to disagree with the results from Figure \ref{sfr}, where no deficiency at that time is apparent. 
If we take into account that the model stellar masses at $z=2$ are slightly high, it seems reasonable to assume the measured SFRs
together with stellar masses according to the MNW13 predictions from abundance matching.
This can explain part of the discrepancy, but it this is not enough
to explain the full difference between observations and models. Interestingly, \citet{kannan} found a similar discrepancy
with observations for their galaxies in a simulation of a cosmological volume.

Another direct comparison to observation can be made for ages. We determine the mean age of populations within
three stellar half-mass radii and compare them to two observational relations in Figure \ref {ages}. \citet{gallazzi} provided
an age-stellar mass relation for a magnitude-limited sample of SDSS galaxies, whereas \citet{thomas} examined a sample
of morphologically selected elliptical galaxies. This figure very clearly reveals the trends discussed above and a division
of our sample in three parts. The trend of ages of all model galaxies with stellar mass is almost flat and thus disagrees 
with the full galaxy population. The Aquarius haloes and the other haloes with masses $\sim 10^{12} M_{\odot}$ are
in reasonable agreement with both observational datasets, which strengthens the conclusions about their well-modeled SFHs.
The massive galaxies with high late SFRs only marginally agree with the youngest observed $M_{\star}\sim 10^{11}M_{\odot}$ galaxies.
The lower mass galaxy models, which include the three haloes with lowest masses and the galaxies undergoing a major merger at $z=0$,
overlap with the observed elliptical galaxies but are more than $1\sigma$ high compared to the
Gallazzi et al. sample, which at these masses is dominated by disks.

In conclusion, our feedback model, calibrated on halo AqC, works well for all haloes with similar and slightly lower masses.
These models show overly high SFRs only at high redshift. In the most massive haloes we studied, we find that SFRs are significantly too high
at low redshift, possibly on account of the lack of a model for AGN feedback. For our lower mass galaxy models we find too low
SFRs at $z<1$, which makes the ages of their stellar populations too high for disk galaxies at these masses.

\section{Morphology and Kinematics}

After studying how the stellar mass assembles in our model galaxies, we now have a closer look at the morphology and 
the kinematics of their stellar components.

\subsection{Structural Properties}

\begin{figure}
\centering
\includegraphics[width=8.5cm]{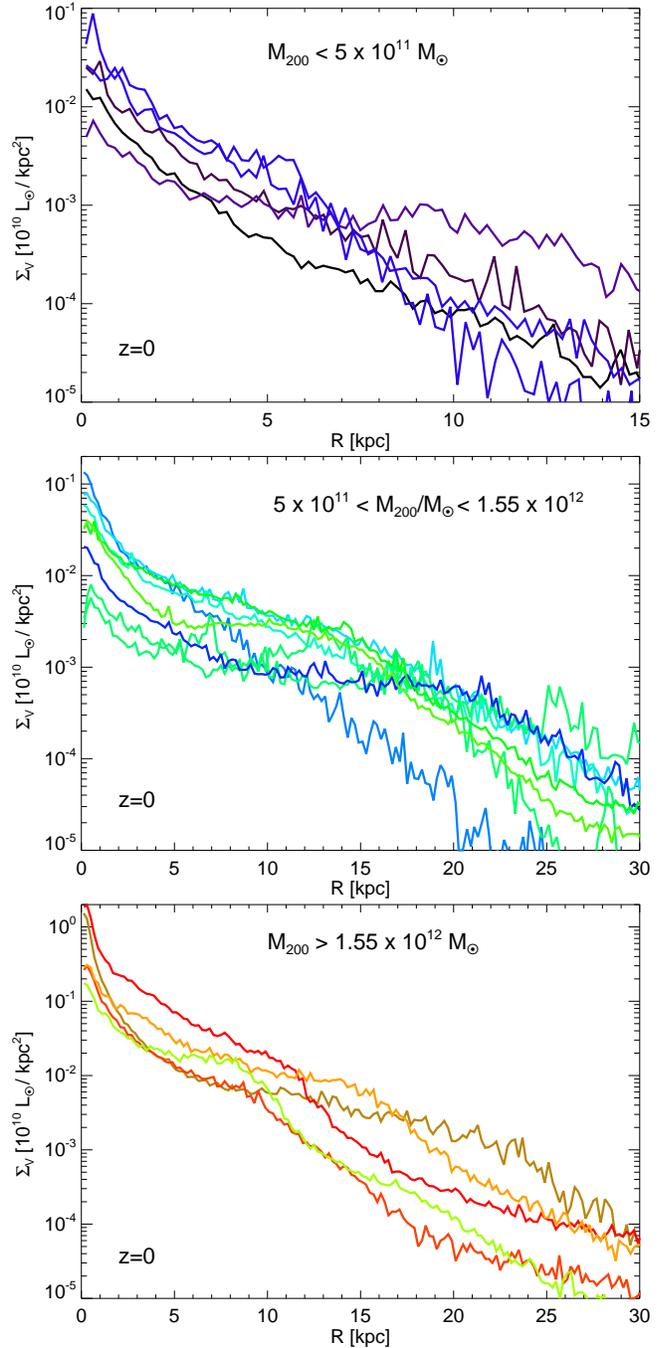}
\caption{Radially averaged $z=0$ face-on $V$-band surface brightness $\Sigma_{V}$ as a function of radius $R$ 
for all our model galaxies. We split the sample into three bins of halo mass: $M_{200} < 5\times 10^{11} M_{\odot}$ (top),
$5\times 10^{11} M_{\odot} < M_{200} < 1.55 \times 10^{12} M_{\odot}$ (middle) and $M_{200} > 1.55\times 10^{12} M_{\odot}$ (bottom).}
\label{ml}
\end{figure}

We start by a visual inspection of galaxy morphologies. We use the models of \citet{bruzual}
to assign $V$-band luminosities to the stellar particles according to their mass, age and metallicity.
We then create edge-on and face-on surface brightness $\Sigma_{V}$ images for the stellar component
without taking obscuration into account. We also create gas mass images
and present the results for eight model galaxies at $z=0$ in Figures \ref{sd2d} and \ref{sd2d2}
to display the variety of galaxy types produced in our simulations.

In Figure \ref{sd2d} we present lower mass galaxies. All of these galaxies do show extended gas disks, as do observed disk
galaxies (e.g. \citealp{things}). Moreover, the gas disks and, with the exception of 6782, also the stellar components
do show warps of varying extent. Warps are also frequently observed in real galaxies \citep{sancisi}.
Considering our full sample of 18 galaxies in the 16 simulated haloes (2 each for 2283 and 0977),
10 galaxies show warps and 11 have gas disks that extend beyond the stellar populations.

The stars of the galaxy in our least massive halo 6782 (first row of Figure \ref{sd2d}) show an ellipsoidal distribution, 
which is only mildly flattened. The gas however lives in an extended, warped disk, which is dominated by two spiral arms.
The gas disk has formed after a major merger at $z=0.4$.
Our model 1646 (second row) also shows elliptical morphology in stars, the visible rotational flattening is however higher. 
Moreover, there are signs of a disk and a misaligned ring. The gas displays a strongly warped, relatively compact disk
with disturbed spiral structure. We will show below that this galaxy features strongly misaligned infall at $z\sim 0.1$
and a counter-rotating stellar population. Misaligned infall and reorientation of disks are frequent phenomena
in our simulations. About half of the $z=0$ disks have experienced reorientation of 40 degrees or more since they started forming.

\begin{figure}
\centering
\includegraphics[width=8cm]{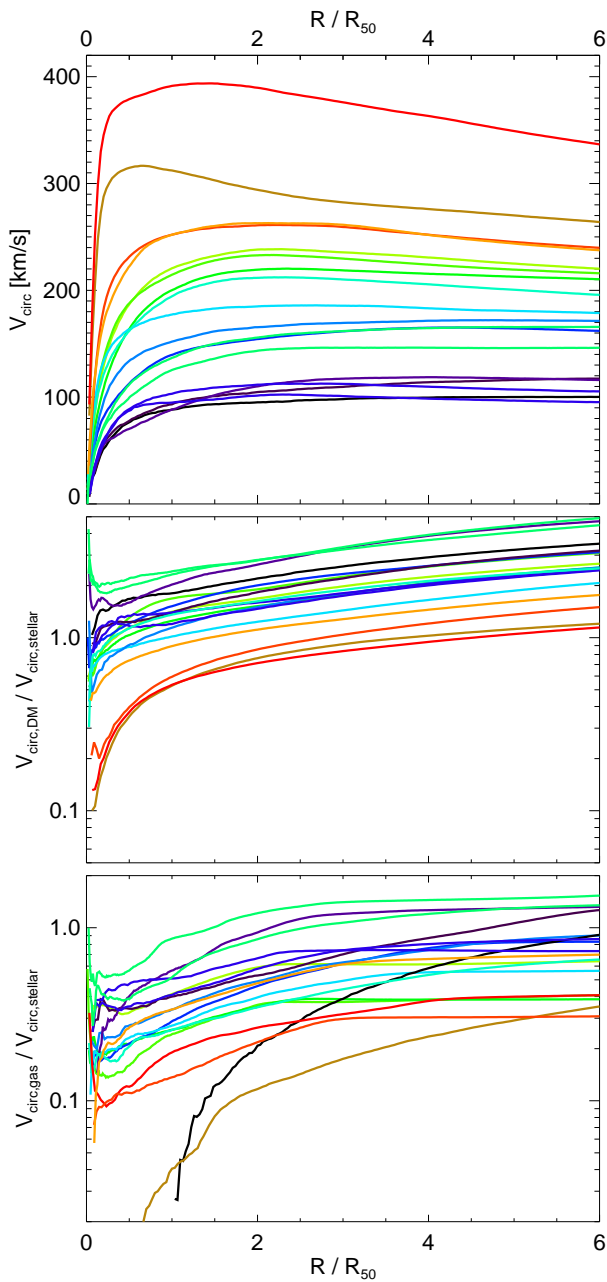}
\caption{Spherically averaged $z=0$ circular velocity $V_{\rm circ}=\sqrt{GM_{\rm tot}(<r)/r}$ for all our models (top panel).
The middle panels shows the fraction of the contributions to $V_{\rm circ}$ by dark matter (DM) and stars $V_{\rm circ, DM}/V_{\rm circ, stellar}$,
where, $V_{\rm circ,i}=\sqrt{GM_{i}(<r)/r}$. The bottom panel depicts $V_{\rm circ, gas}/V_{\rm circ, stellar}$.}
\label{vel}
\end{figure}

Halo AqB (third row) hosts a very thin, extended stellar disk galaxy with flocculent spiral structure 
and a prominent bulge. The gas forms a very extended, slightly warped disk with spiral structure as in the stellar component.
One of the disks in the major merger halo 0977 (fourth row) shows similarities to the one in AqB in terms of spiral structure and central bulge.
However signs of the interaction with its neighbour are clearly visible. 
The system shows elongated tidal features and the stellar disk has been considerably thickened. The edge-on views reveal that
the SF regions in the spiral arms do not lie in a well-defined plane as in AqB but in projection superpose to create a thick disk.

In Figure \ref{sd2d2} we display four galaxies with higher masses. 0959 and 0616 show overly high SFRs at $z=0$, whereas AqD
and 0664 have reasonable $z=0$ masses. For these galaxies, only 0959 (second row) shows an extended, warped gas disk.
It also displays a prominent stellar bar and a multi-arm spiral pattern. Only three of our 18 $z=0$ galaxies show clear signatures
of bars, though more galaxies show bar signatures during their evolution. The $z=0$ rate of bars is thus lower than in real galaxies \citep{eskridge}.
The model galaxy in halo 0664 (third row) has a low gas fraction with gas being confined to a compact, un-warped disk
The edge-on stellar morphology shows a disk and an underlying spheroidal distribution, the face-on image shows
no distinct spiral structure. The morphology is thus reminiscent of S0 galaxies.

In the next subsection, we will show that AqD has the highest kinematic disk fraction of all our model galaxies.
The first row of Figure \ref{sd2d2} reveals a large stellar disk with a small bulge and faint spiral structure.
The gas lives in a thin, un-warped disk, which is less extended than its stellar counterpart. 
The morphology of the galaxy in halo 0616 (fourth row) is clearly also disky and features spiral arms. However, the disk appears significantly thicker
and more compact. The gas disk is also compact and thick explaining the high SFRs. Moreover, the
gas distribution extends vertically beyond the disk. We find it to be inflowing as part of a galactic fountain.

Disk galaxies have long been known to show roughly exponential surface density profiles with central bulge concentrations \citep{devau}.
In the recent years it has become clear that most of these exponentials actually consist of two or more parts which themselves
are approximately exponential. Most of these breaks (or truncations) in profiles are down-bending, however a significant fraction
of galaxies has up-bending profiles (see e.g. \citealp{s4g}).

As the majority of our galaxies are disky their profiles should show similar shapes, which indeed they do.
In Figure \ref{ml} we plot radially averaged face-on $V$-band surface brightness profiles for all our model galaxies
and split them into three panels according to halo mass (note the different scales). We find several interesting details:
The lower mass galaxies are less extended than higher mass galaxies, as observed \citep{courteau}.
Also most galaxies do show (almost) exponential parts in their profiles. As in real disk galaxies, the central excess
is more pronounced for higher mass galaxies and there are low mass galaxies showing no or hardly any central upturn.
Four out of 18 galaxies have profiles that can be fit with a single Sersic profile with $n \ge 1$, one of which
could be classified as 'pure-exponential'.

\begin{figure*}
\centering
\includegraphics[width=16cm]{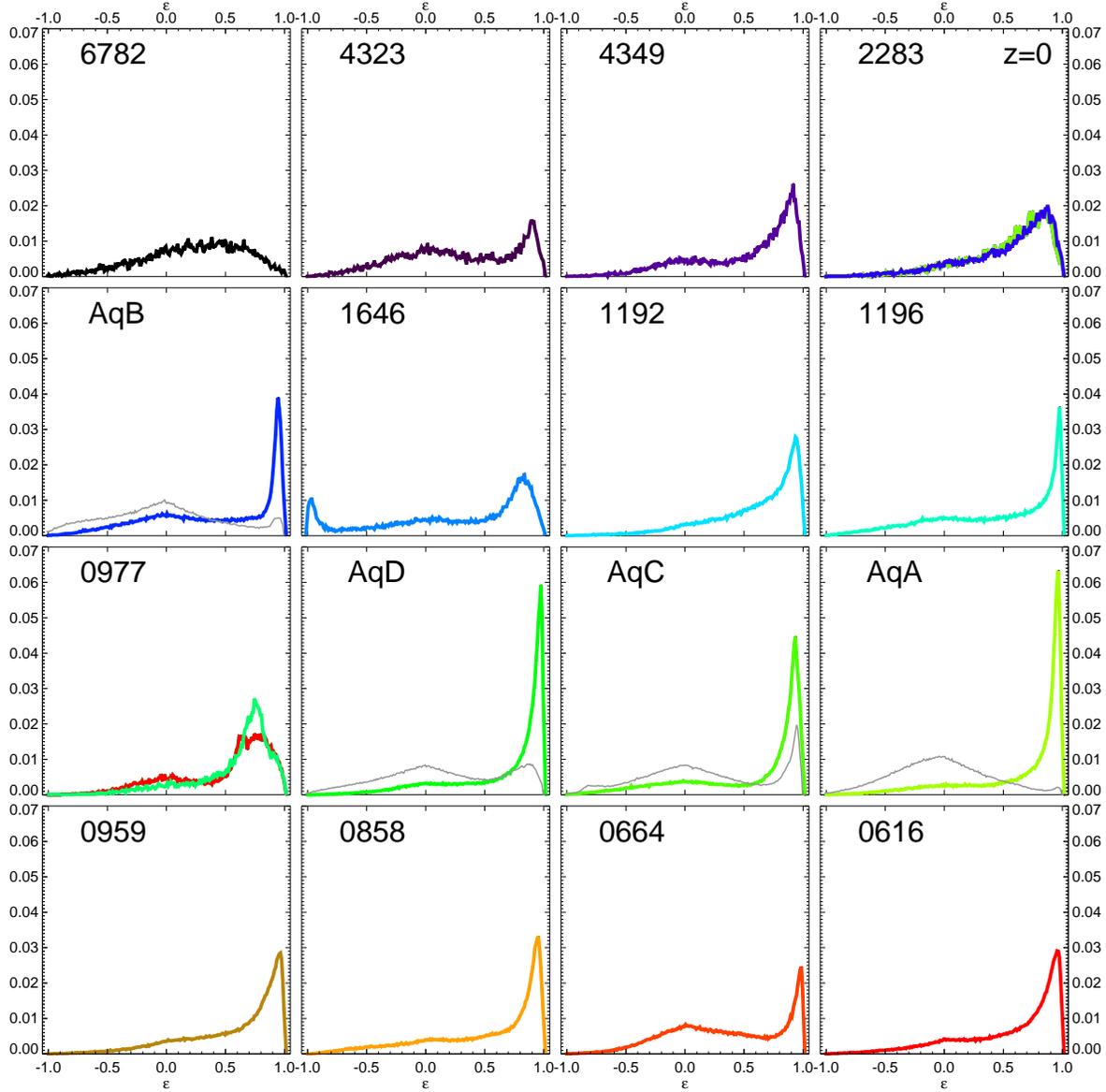}
\caption{Distributions of circularity $\epsilon=j_{\rm z}/j_{\rm circ}(E)$
for our model galaxies at $z=0$. The panels for the mergers 2283 and 0977 contain distributions
for both galaxies. For 2283 the two galaxies show almost indistinguishable distributions.
For haloes AqA, AqB, AqC and AqD we overplot the $\epsilon$-distributions for the corresponding models of CS09 in grey.}
\label{epsilon}
\end{figure*}

Out of the galaxies with a central component, one has a single exponential outer component.
In the remaining model galaxies, the disk components feature breaks. Nine of them are bulge+disk+down-bending profiles, 
but four of our massive galaxies (lower panel) show up-turning features in their outskirts as well.
So from a qualitative point of view, the distribution of surface brightness profiles
is in reasonable agreement with observations.

Cosmological simulations of galaxy formation have long suffered from an over-concentration of baryons in the centers of haloes
\citep{navarrobenz}. This is reflected by circular velocity $V_{\rm circ}=\sqrt{GM(<r)/r}$ curves that are strongly centrally peaked
(see e.g. some models in CS12). Observed gas rotation curves of galaxies are however fairly flat
or rising outwards in low mass disk galaxies \citep{deblok}.
To reproduce this observation, it has been shown that feedback models have to selectively remove low angular momentum
gas (e.g. \citealp{brook}).

\begin{figure*}
\centering
\includegraphics[width=18cm]{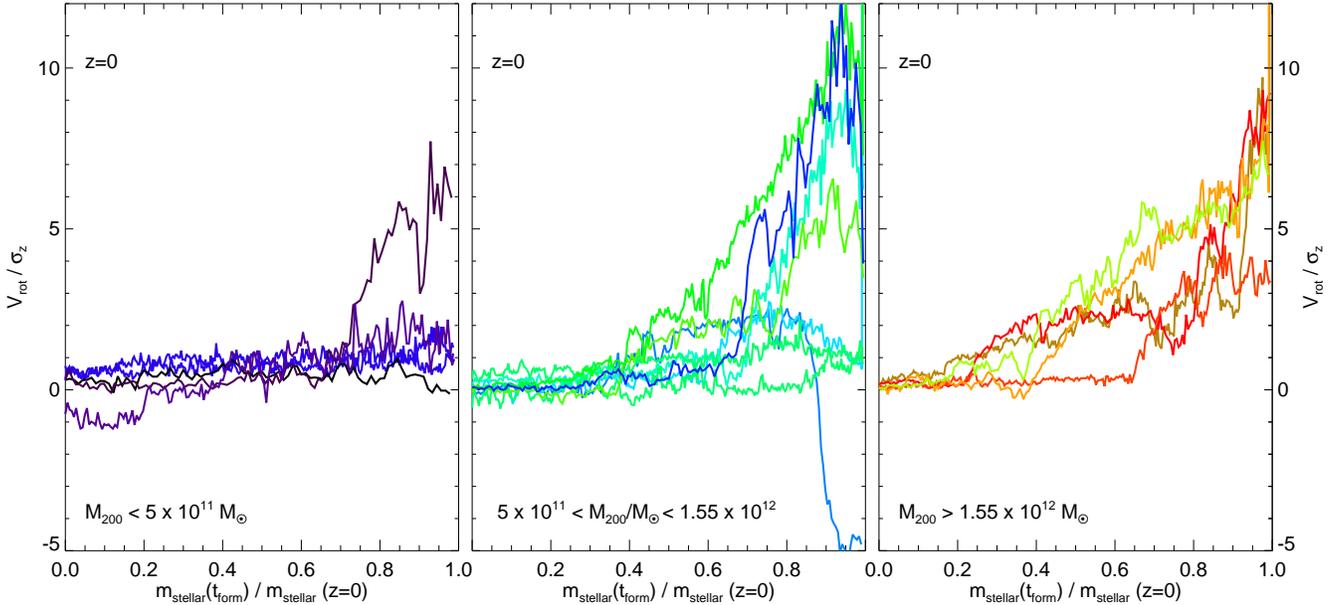}
\caption{Rotation to dispersion ratios $V_{\rm rot}/\sigma_z$ at $z=0$ as a function of the fraction of stellar galactic mass that has formed
until the formation time $t_{\rm form}$ of a star particle relative to the stellar galactic mass at $z=0$.
Our model galaxies are split into three bins in halo mass: $M_{200} < 5\times 10^{11} M_{\odot}$ (left),
$5\times 10^{11} M_{\odot} < M_{200} < 1.55 \times 10^{12} M_{\odot}$ (middle) and $M_{200} > 1.55\times 10^{12} M_{\odot}$ (right).}
\label{vs}
\end{figure*}

In the top panel of Figure \ref{vel} we plot circular velocity curves for all our model galaxies as a function
of the radius normalized to the stellar half mass radius $R/R_{50}$. Only the
two galaxies in 0959 and 0616, which show a strong overproduction of stars at low $z$ and feature compact, thick gas-rich disks
show central peaks in $V_{\rm circ}$, but to a less severe extent than e.g. several models in CS12.
The other galaxies indeed show very flat circular velocity curves with detailed shapes depending on the details
of their formation histories. In general, our feedback mechanism is thus capable of producing realistic
central mass distributions.

In the other panels of Figure \ref{vel} we show the fractions of the contributions of dark matter, stars and gas to the $V_{\rm circ}$ curves.
Clearly the centrally peaked models are dominated by stars. It is interesting to note that there is a continuous
transition in our galaxies from low-mass galaxies for which at all radii the gravitational potential is dominated
by dark matter to more massive galaxies, which are star-dominated in the centers and for which dark matter
becomes dominant at a radius that increases with galaxy mass. As far as gas is concerned, the baryonic contribution
is dominated in the centers by stars for all galaxies. However for low-mass, more gas-rich galaxies, the gas contribution
surpasses the stellar contribution at the outskirts of the galaxies.

In summary, our model galaxies display a range of realistic morphologies in gas and stars. Unlike in many previous simulations
they are not too centrally concentrated and show no significant disagreement with observed galaxies in terms
of radial surface density profiles and circular velocity curves.

\subsection{Circularity distributions}

We now want to connect morphology to stellar kinematics.
Circularity $\epsilon=j_{\rm z}/j_{\rm circ}(E)$ distributions \citep{abadi} are a tool that have been 
widely used to quantify stellar disks in simulations. Here $j_{\rm z}$ is the $z$-component of the specific angular momentum
of a stellar particle in a cylindrical coordinate system with the disk axis as symmetry axis and $j_{\rm circ}(E)$
is the specific angular momentum of a particle of energy $E$ on a circular orbit.
Particles on circular orbits thus show $\epsilon=1$ and a thin disk results in a peak close to this value. A bulge will
results in a peak around $\epsilon=0$. 
We have computed such distributions for all of our models and present them in Figure \ref{epsilon}.

We notice, that apart from 6782, the lowest mass halo, all our models show distinct disk peaks. 6782 shows an asymmetric distribution
indicating some rotational support, but no stellar disk as we had observed in the first row of Figure \ref{sd2d}. 
The spheroid results from a major merger at $z\sim 0.4$ and a lack of SF afterward. 4323 shows a small disk and a dominant 
bulge, whereas 4349 shows a broadened disk peak indicating a rather thick disk. Considering that all these haloes have overly low SFRs
in the low-$z$ era of disk formation, the lack of dominant thin disks is not surprising.

For the haloes with $z=0$ major mergers, 2283 and 0977, we plot distributions for both galaxies. In each case they show very broad disk peaks,
indicating that these are mergers of disk galaxies with similar masses. The galaxies are still separated but the gravitational interaction has 
already considerably heated the stellar disks. The thickening of one of the galaxies in 0977 is clearly visible in the fourth row of Figure \ref{sd2d}.

The most peculiar distribution of circularities is exhibited by 1646. It consists of two disks with opposite directions of rotation
that live on top of each other. The co-rotating disk is more massive, but is also considerably thicker than the young and relatively thin
counter-rotating disk. Additionally, a distinct bulge peak is visible. This composition explains the disturbed and more elliptical
morphology visible in the second row of Figure \ref{sd2d}. The model galaxy in halo 1192 shows a disk peak with a wide tail 
to lower values of $\epsilon$, which is connected to a bar present throughout the evolution of the disk.

Among the most massive haloes, 0664 has the lowest low-$z$ SFR and a correspondingly small disk peak,
which has formed after a merger at $z\sim 0.7$. The circularity distribution
is dominated by a broad bulge peak resulting from the merger. This explains the S0 like morphology seen in Figure \ref{sd2d2}.
0959, 0858 and 0616 show dominant disk peaks with tails to lower values of $\epsilon$ indicating rather thick
disks. The $\epsilon$ distribution of 0959 is shaped by the presence of a bar (see Figure \ref{sd2d2}, whereas the disks in 0858 and 0616
are heated by minor mergers occurring throughout their evolutions. The bottom row of Figure \ref{sd2d} correspondingly shows a compact 
and thick stellar disk in 0616. 

\begin{figure*}
\centering
\includegraphics[width=5.6cm]{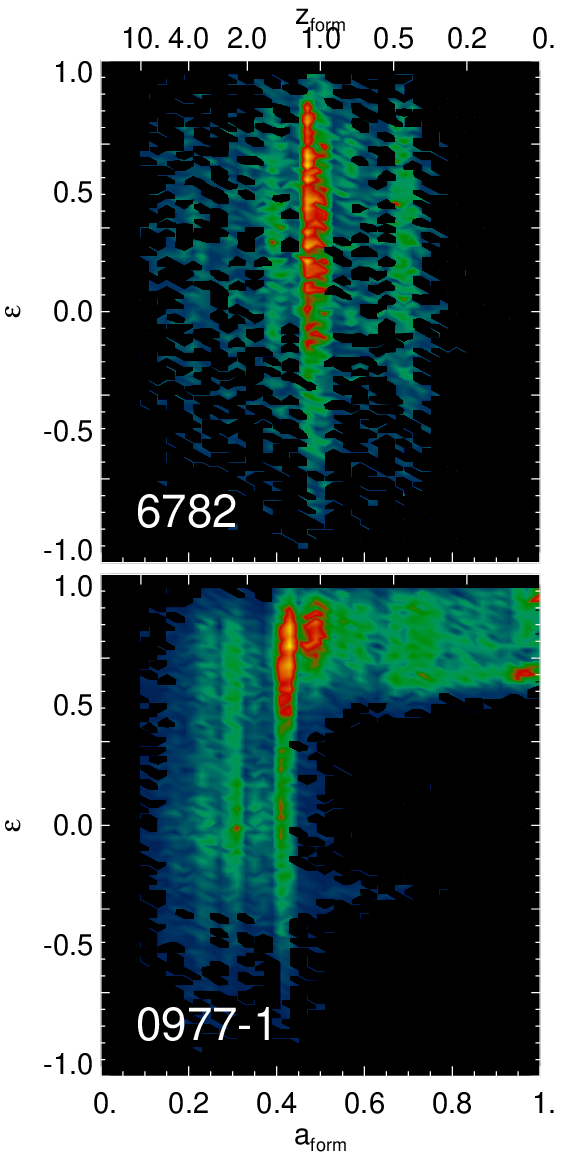}\includegraphics[width=4.8cm]{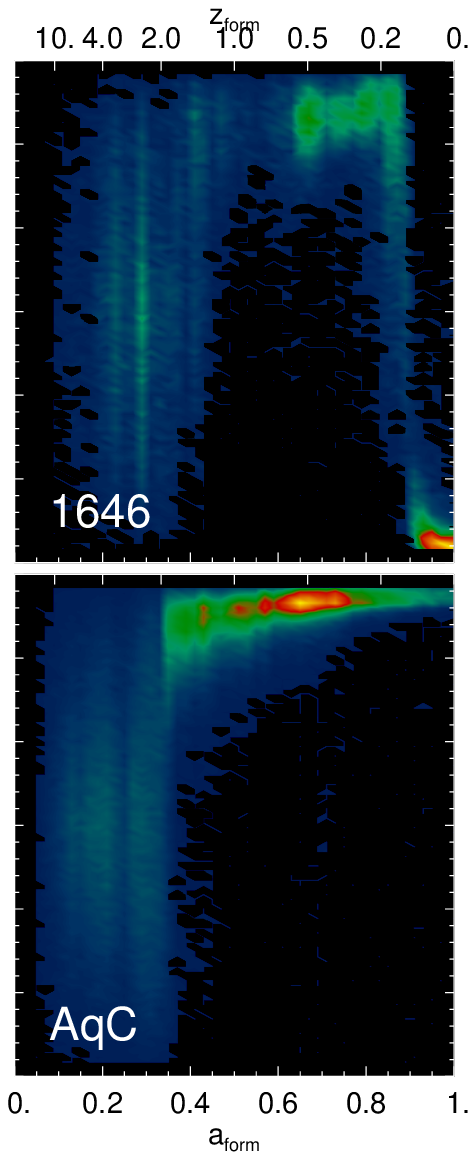}\includegraphics[width=7.2cm]{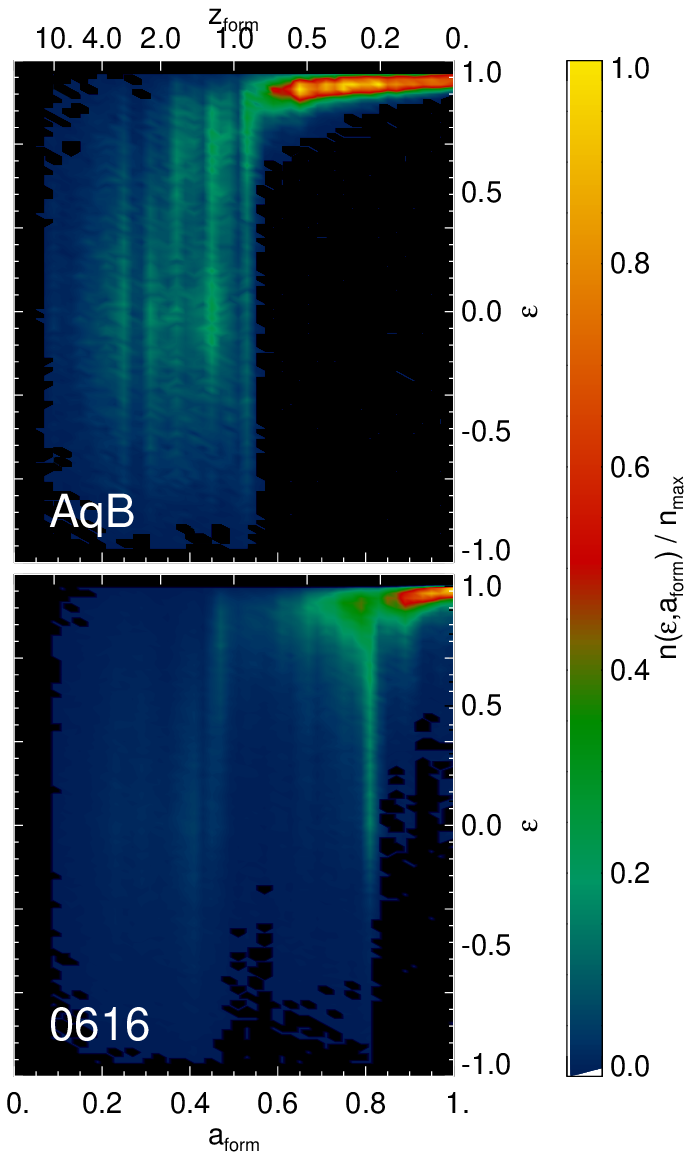}
\caption{The normalized number densities of the stars in six of our galaxy models at $z=0$ over the plane 
of circularity $\epsilon$ vs. formation scale factor $a_{\rm form}$, $n(\epsilon,a_{\rm form}/n_{\rm max})$.
The models are from top to bottom: 6782, 1646, AqB, one of the two merging galaxies
in 0977, AqC and 0616.}
\label{epsage}
\end{figure*}

The most dominant and thinnest disk peaks indicating high fractions of thin, star forming disks can be found in the panels for the Aquarius
haloes. Of the haloes from \citet{oser} only 1196 shows a comparably thin disk peak, which however only forms after a merger at $z\sim 0.7$.
This is not surprising as the Aquarius haloes were selected to be prime candidates to host such galaxies.
However, CS09 were not able to completely confirm this conception in their study of all haloes, as their highest kinematic disk fractions
were $\sim 25 \%$. Higher disk fractions were found for different codes for AqC in CS12,
but these galaxies did not compare well to observed disk galaxies. \citet{okamoto} presented the best
Aquarius cosmological disk models so far for haloes AqC and AqD, their circularity distributions are however less disk dominated than ours.
Considering the idealized, semi-cosmological models in AqA and AqC from AW13, our cosmological simulations produce disk fractions
similar to the reference model ARef, but not as good as the best models presented there. This is not surprising, as these models ignore the evolution
at $z>1.3$ when destructive mergers occur.

To illustrate the change in circularity distributions between the models of CS09 and our models we over-plot in Fig. \ref{epsilon}
the corresponding curves for their models of AqA, AqB, AqC and AqD in grey. 
\footnote{Note that the circularity distributions presented in Figure 3 of CS09 were
based on a different definition of circularity, $\epsilon_{\rm V} = j_{z}/(rV_{\rm circ}(r))$ with $V_{\rm circ}=\sqrt{GM(<r)/r}$.
Comparing Figure 3 of CS09 to Figure 2 of \citet{tissera}, who used the same definition as we did and analyzed the models of CS09, illustrates
how the definition of circularity affects the shape of the circularity distribution.}
For AqA they found hardly any disk and correspondingly the 
difference between their bulge-dominated and our disk-dominated model is very prominent. For AqC they found their highest disk fraction,
but the increase in disk-fraction by a factor of almost 3 is still clearly visible. For AqB and AqD, the increase in disk fraction
is also striking. 
This comparison clearly favours the use of very efficient feedback recipes in simulations. We however note that
more conservative approaches to feedback models have also been successful in producing distinct disk peaks in circularity
distributions (see e.g. \citealp{governato, few}).

In summary, our models show a large variety of kinematic properties ranging from rotating spheroids, bulge-dominated disks
and counter-rotating components over disks thickened by misaligned infall, bars and mergers to compact, massive disks and dominant
thin disks.

\subsection{Disk Fractions}

\begin{figure*}
\centering
\includegraphics[width=17.5cm]{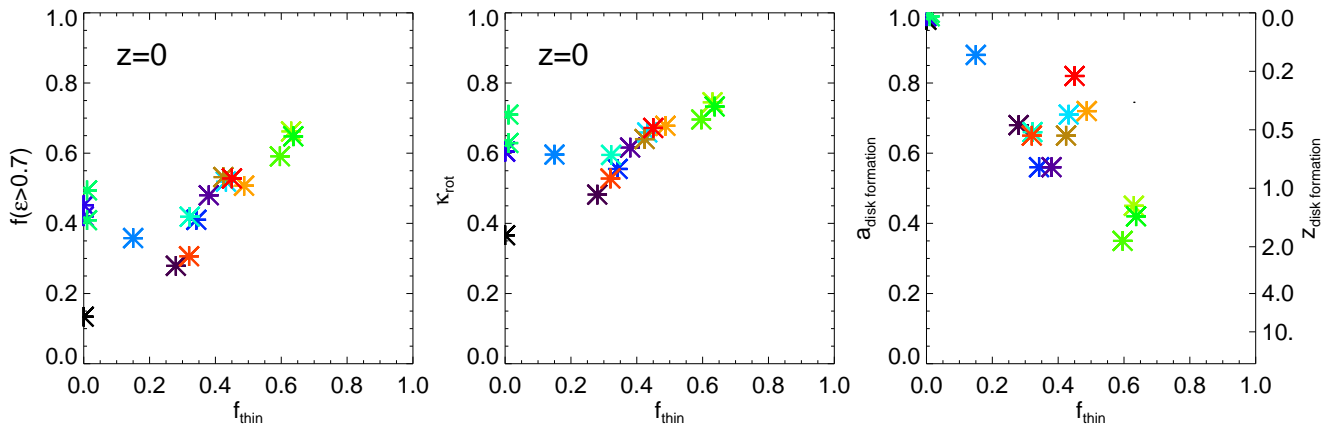}
\caption{An overview of the disk fractions at $z=0$ in our model galaxies. In all panels we compare to our definition of 
thin disk fraction $f_{\rm thin}$ given by the mass fraction in the 'thin disk area' of the $\epsilon$-$a$ plane (see Fig. \ref{epsage}).
The left panel shows $f(\epsilon>0.7)$, the fraction of stars with circularity $\epsilon>0.7$,
the middle panel the fraction of the kinetic energy in rotation $\kappa=E_{\rm rot}/E_{\rm kin}$ as defined by \citet{sales}
and the right panel the scale factor when disk formation starts, $a_{\rm disk formation}$.}
\label{diskfraction}
\end{figure*}

Disk fractions in simulations are sometimes quantified by mock photometric surface brightness decompositions (see e.g. \citealp{agertz,eris}).
\citet{cs10} showed that this way of estimation yields significantly higher disk fractions than kinematic decompositions.
In this work, we focus on kinematic disk fractions way as they constitute a more fundamental characteristic of the stellar population.

To have a direct comparison to the models presented in CS12 we have calculated
a distribution of circularities $\epsilon_V$ according to their definition based on the local circular velocity of a stellar particle.
We then determined the fraction $f_{\epsilon}=\epsilon_V>0.8$. For AqC we find $f_{\epsilon}=0.55$. All models in CS12 showed
$f_{\epsilon}<0.55$, with the highest values occurring in galaxies with centrally peaked rotation curves. Moreover, our $\epsilon_V$ distribution for 
AqC shows a higher ratio of disk peak to bulge peak than those galaxies. We thus conclude that our model
produces a better disk than any of those models.

In AW13 we argued that rotation-to-dispersion ratios are a good additional tool to characterize disks. 
We sort all stars in a galaxy as a function of age and bin them in equal-mass bins.
We define a disk plane perpendicular to the total stellar angular momentum and calculate $V_{\rm{rot}}$ as the mean tangential velocity
and $\sigma_z$ as the rms velocity in the direction of the rotation axis for all stars in a certain age bin.
In Figure \ref{vs}, we plot the ratio $V_{\rm rot}/\sigma_z$ as a function of the fraction of the $z=0$ stellar mass that has formed 
up to the time in consideration. We again split the models into three bins according to halo mass and include two galaxies per
major merger halo.

The idealized disk models from AW13 showed a monotonic decrease of $V_{\rm rot}/\sigma_z$ with increasing age. However these models
did not include mergers, which for all haloes occur at high $z$. We thus expect an additional component 
with $V_{\rm rot}/\sigma_z\sim 0$ for the oldest ages.
Moreover, the idealized models showed that, at the resolution in consideration, values of $V_{\rm rot}/\sigma_z>>10$ as they are observed
for the Milky Way thin disk (e.g. \citealp{ab}) are prevented by the coarse resolution and the specifics of the modeling of ISM physics
(see also \citealp{house}). In general populations of stars with $V_{\rm rot}/\sigma_z>2$ can be interpreted as disky and the 
plot can thus be used to estimate disk fractions.

Significant counter-rotating components are only found for the 15 \% youngest stars in 1646 as discussed above and for the mildly (counter-)rotationally
supported oldest bulge component of 4323. All models show dispersion-dominated bulge populations for their oldest components, as expected.
For the low mass haloes only 4349 shows a significant 'thin' component with $V_{\rm rot}/\sigma_z>5$. The merger-thickened disks
in haloes 2283 and 0977 show $V_{\rm rot}/\sigma_z<2$ for all populations and the disk populations of 1192 and 1646, which suffer from 
a bar or misaligned infall, all show $V_{\rm rot}/\sigma_z<5$. So, as suggested in AW13, misaligned infall or reorientation of the disk plane
heats existing stellar populations and prevents the formation of thin disks. 

All other haloes show distinct disk populations as discussed for Figure \ref{epsilon}. The highest ratios $V_{\rm rot}/\sigma_z\sim 10$
are displayed by the disks in AqB and AqD, but the massively star forming disks in the most massive haloes also reach 
$V_{\rm rot}/\sigma_z\sim 8$. If disks are present their $V_{\rm rot}/\sigma_z$ ratios decrease almost monotonically with age,
a sign for continuous disk heating (e.g. \citealp{jenkins}), as discussed in AW13. The disks with the highest fractions of populations
with $V_{\rm rot}/\sigma_z>2$ are AqA and AqD with fractions similar to 60 \%.

To gain a better understanding for the formation times of stellar disk components 
in our simulations we present in Figure \ref{epsage} the distribution of stellar 
mass in the plane of circularity $\epsilon$ vs. formation scale factor $a_{\rm form}$. We present six different models to give an overview
of how different evolution histories show up in this plot.
The SFH of 6782 is bursty and stops at $z\sim 0.4$, when a major merger occurs. 
Stars of all ages show a similar distribution of circularities, which characterizes
an elliptical galaxy with small net rotation.

A merger is still ongoing in halo 0977. Here we see two-stage formation. The stars that formed at $z>1.3$ show a bulge distribution similar to 
6782, whereas all younger stars live in a disk, which has been disturbed by the ongoing merger. The galaxy in 1646 shows three-stage formation.
Again stars that formed at $z>1.3$ live in a bulge, whereas stars that formed at $1.3>z>0.15$ make up a thick disk. In this case
the disk was heated by a drastic change in the angular momentum orientation of the infalling gas starting at $z\sim 0.25$.
This reorientation results in the formation of a thinner counter-rotating disk.

In 0616 the picture is more complicated. Stars at all formation times $a<0.8$ show a broad bulge distribution. At $a=0.8$ a last destructive merger
occurs. However there is an older disk population, formed at $0.6<a<0.8$, that has survived thickened by the merger.
A thinner disks forms after $a=0.8$, but there are also young stars with thick disk circularities, which are connected to ongoing
interactions with satellite galaxies after $a=0.8$.

This is not the case for the disks in AqB and AqC. For both a dominant disk population is visible
at high $\epsilon$ after some well-defined time when formation starts, which for AqB is a destructive merger at $z\sim1$.
In AqC disk settling starts around $z=2$ when mergers stop disturbing the galaxy.
The disk populations widen in $\epsilon$ for older ages, again indicating continuous disk heating.

These $\epsilon$ vs. $a$ plots provide us with a good procedure to define disk fractions. We can define a 'disk area' in this
diagram, which for AqB is clearly given by $a>0.55$ and $\epsilon >0.8$. The lower border for $\epsilon$ is not well-defined,
however $\epsilon\sim0.75$ works well for all examples. We hereby also define a time when disk formation starts.
With this definition, 6782 and 0977-1 have no disks, the latter as the stellar system has been significantly heated by the ongoing merger.
We therefore refer to this definition of disk fraction as 'thin disk fraction' $f_{\rm thin}$, but caution that observed thick disks would not necessarily
be excluded by this method.

For 1646, the counter-rotating disk is counted as the thin disk and for 0616 only the disk that formed at $a>0.8$ is included.
We use this procedure to determine $f_{\rm thin}$ for all model galaxies and use the values in Figure \ref{diskfraction}.
In the right panel we plot disk formation time $a_{\rm form}$ against $f_{\rm thin}$.
The values for $f_{\rm thin}$ range between 0.15 (the counter-rotating disk in 1646) and 0.64 (AqA and AqD). We clearly see
a correlation in the sense that younger disks have smaller mass fractions, which is not surprising. It is however interesting
to have qualitative predictions for the connection of the time of the last destructive event and the final disk fraction.
It is also interesting that the best disks in the Aquarius haloes AqA, AqC and AqD separate from the rest of the population
indicating that their selection criteria indeed yielded haloes which host dominant disk galaxies.

\begin{figure}
\centering
\hspace{-0.7cm}\includegraphics[width=9cm]{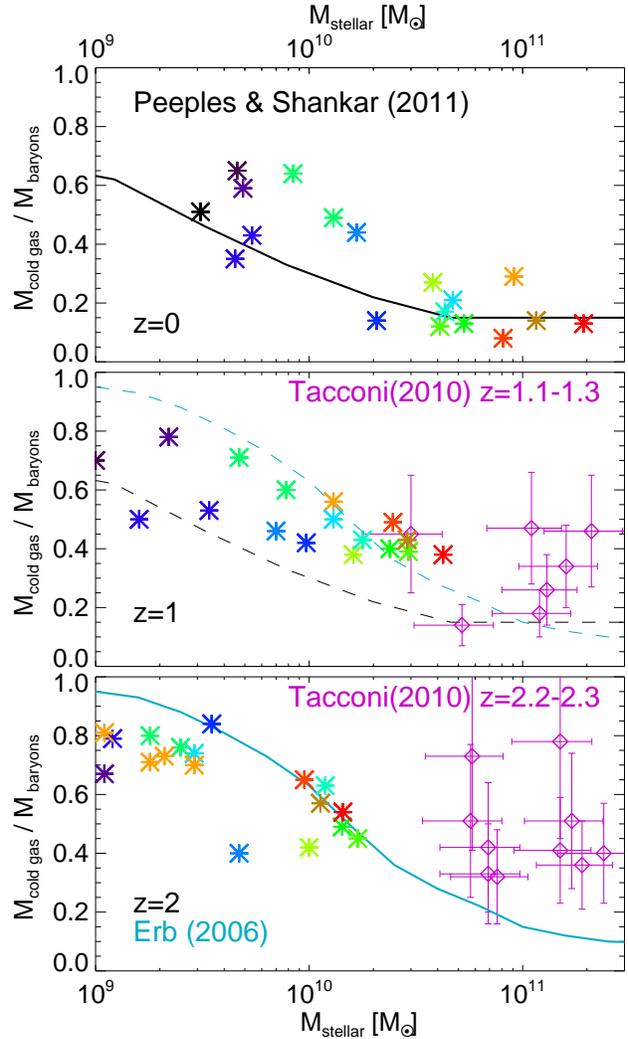}
\caption{The fraction of the mass in gas with $T<10^4\;\rm{K}$ to total baryonic mass within 20 kpc vs. 
the galactic stellar mass $M_{\rm stellar}$ at redshifts, $z=0,1$ and $2$. 
Overplotted are the observed mean $z=0$ relation presented by \citet{peeples}, an approximate
relation for $z=2$ obtained from the results of \citet{erb} and data from \citet{tacconi}.}
\label{gasf}
\end{figure}

We also compare to previously used definitions to quantify disks. CS12 used all stars with $\epsilon_V>0.8$. We find that for our definition
of the circularity $\epsilon>0.7$ gives similar results. From the left panel in Figure \ref{diskfraction} we see a clear correlation
between the two disk fractions. However a simple $\epsilon$ cut assigns a disk fraction to the thick merging systems as well as to the
elliptical galaxy 6782. \citet{sales} used the fraction of kinetic energy in rotation $\kappa=E_{\rm rot}/E_{\rm kin}$ as an indicator for
diskiness. $\kappa$ increases with increasing $f_{\rm thin}$ except for ongoing mergers.

In conclusion, we have shown how rotation-to-dispersion ratios and the distribution of stars in the plane of circularity
vs. formation time can be used to connect the kinematics and the formation history of model galaxies. Our galaxies are not
as dynamically cold as real galaxies due to numerical issues. However, they realistically show that the oldest components
of galaxies are dispersion dominated and that rotational support increases with decreasing age of stars as observed in disk galaxies.
We have shown that the model disk fractions in our Aquarius galaxies are higher than in any previously published simulation
for these haloes. The disk fraction in our simulations depends mostly on the time since the last destructive event, be it
merger or misaligned infall.

\section{Scaling Relations}

We continue our study of the galaxy population constituted by our models and have a look at how they compare to observed
relations involving stellar and gas masses, rotation velocities, sizes and metallicities.

\subsection{Gas fractions}

Observed gas fractions in $z=0$ galaxies show a clear trend of higher gas fractions at lower galaxy masses (\citealp{haynes},
see \citealp{peeples} for a recent compilation of data). At $z=1-2$, \citet{tacconi} using measurements of CO 
showed that gas fractions at a given stellar galactic mass are on average significantly higher.
They however analyzed galaxies more massive than the ones in our simulations. \citet{erb} provided gas fractions for lower mass $z\sim2$ galaxies by
estimating the gas mass applying the local Kennicutt-Schmitt relation to SFR measurements. At higher masses, they
found significantly lower gas fractions than those measured by \citet{tacconi}.

These observations provide an important test for our models of star formation
and feedback in the context of cosmological gas accretion. Star formation uses up gas, whereas feedback can prevent gas from cooling 
and thus forming stars or eject gas from galaxies. Ejected gas can leave the halo if energetic enough or come back in galactic fountains.
Clearly inappropriate modeling of these processes can thus be identified from such a relation.

In Figure \ref{gasf} we attempt such a comparison. We determine the mass of gas with $T<10^4 \; \rm K$ within 20 kpc and use it
to determine the fraction of cold gas to total baryonic mass in the galaxy. At $z=0$, our simulations reproduce the observed decrease
in gas fraction with increasing stellar mass which becomes shallow or potentially flat at masses $M_{\star}>10^{10.5}\;M_{\odot}$.
For the shallow part of the relation, the gas fractions in our model galaxies scatter around $\sim15$ \%, as in real galaxies.
For the lower mass galaxies, gas fractions can be as high as 65 \% and the low-mass population in total shows a higher
mean than observed. The low-mass galaxies that lie significantly above the relation all live in haloes which show
low stellar masses in the $M_{\rm stellar}-M_{200}$ relation discussed in Figure \ref{moster}. Thus for these galaxies
the low SFRs at low-$z$ lead to an underproduction of stars despite a significant reservoir of gas in the model galaxies.
This inefficient SF then leads to overly high gas fractions at $z=0$.

As in real galaxies, the gas fractions in our model galaxies at fixed $M_{\rm stellar}$ steadily increase when going to redshifts
$z=1$ and $2$. At $z=1$, the gas fractions for $M_{\rm stellar}<10^{10}M_{\odot}$ are intermediate to the $z=0$ and $z=$ 
observational constraints. For $M_{\rm stellar}>10^{10}M_{\odot}$, the gas fractions agree well with the observations
of \citet{tacconi}. At $z=2$, our models lie on top of the observational relation provided by \citet{erb}, with only a slight 
discrepancy towards low gas fractions for the lowest stellar masses, consistent with the fact that these galaxies show too high stellar
masses at these times. Considering the gas fractions measured by \citet{tacconi} for higher mass galaxies, our gas fraction 
in higher mass galaxies at $z=2$
appears too low. \citet{nara} recently reported a similar discrepancy for the simulations of \citet{dave}. They argue that
this discrepancy arises from an overestimation of gas fractions based on incorrect CO-to-$\rm{H_{2}}$ conversion factors.

In summary, the evolution of gas fractions in our model galaxies reproduce observed trends well. Deviations from observational data
can be linked to deviations from the predicted SFH and uncertainties in high-$z$ observations.

\subsection{Sizes}

\begin{figure}
\centering
\hspace{-0.7cm}\includegraphics[width=9cm]{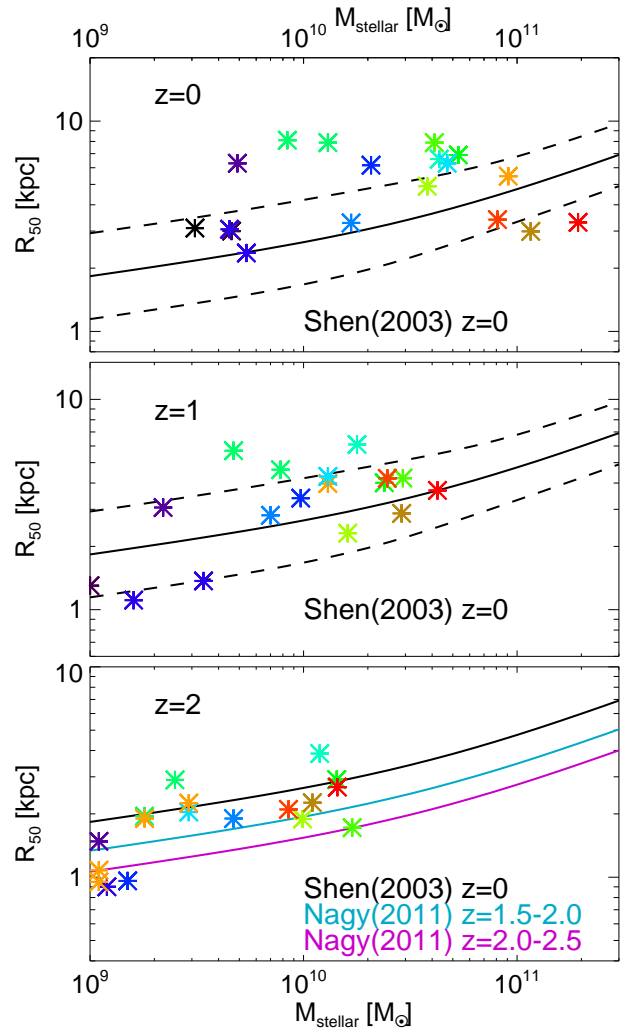}
\caption{Stellar half-mass radius $R_{50}$ vs. galactic stellar mass $M_{\rm stellar}$ at redshifts $z=0,1$ and $2$. 
Overplotted in  all panels is the observed relation for late-type SDSS galaxies by
\citet{shen03} for $z=0$ with 1$\sigma$ scatter denoted by dashed lines.
\citet{nagy} find that galaxies at $z=1.5-2.0$
are typically 27 \% and galaxies at $z=2.0-2.5$ typically 42 \% smaller than galaxies of the same mass at $z=0$.
We overplot the corresponding lines in the $z=2$ panel.}
\label{sizes}
\end{figure}

Another important test for galaxy formation models is the size distribution of galaxies. Galaxy sizes depend on the distribution of
baryonic angular momentum and reproducing this distribution is crucial for achieving correct morphologies, and thus
also SFRs, gas fractions etc. Cosmological simulations have long suffered from an over-concentration of low angular momentum
material in the centers of haloes \citep{navarrobenz}. It has been shown that in order to reproduce observed galaxy sizes
the invoked feedback mechanism has to preferentially remove gas with low angular momentum \citep{dutton}. Thus feedback modeling is
effectively tested by galaxy sizes.

In Figure \ref{sizes} we attempt a comparison of our galaxy model sizes with observations by \citet{shen} 
for SDSS late-type galaxies and by \citet{nagy} for $z=1.5-3.0$ star-forming galaxies.
The half-mass radius $R_{50}$ of observed galaxies at all redshifts increases on average
with galaxy stellar mass $M_{\rm stellar}$. \citet{nagy} find that galaxy sizes at a given $M_{\rm stellar}$ increase
with time, being $\sim 41\%$ smaller than local galaxies at $z=2.0-2.5$ and $\sim 27\%$ smaller at $z=1.5-2.0$.
This suggests that around $z=1$ galaxies start to lie on the local relation (see also \citealp{barden}).

For our model galaxies, we also detect such an increase of $R_{50}$ with time.
At $z=2$ the sizes scatter around values similar to those found by Nagy at $z=1.5-2.0$, whereas
at $z=1$ and at $z=0$ they scatter around the $z=0$ relation. At $z=1-2$ the increase of $R_{50}$ with stellar mass
also agrees well with observations. At $z=0$ our models lie on an overly flat relation.
We already showed that the most massive haloes in our sample at $z=0$ host compact, overly star-forming galaxies. Here these galaxies
explicitly appear as too small. On the other hand, our sample also includes two major mergers. The two galaxies in halo 0977
show elongated features due to their gravitational interactions and thus are most extended relative to the observations.
Ignoring these model galaxies yields a relation with reduced scatter and a clear trend of increasing size with increasing mass, which
is, however, still shallower than observed.

Considering the connection of sizes and feedback modelling we note
that our prescriptions are capable of reproducing observed galaxy sizes well at $z \ge 1$, but
tend to be too strong at late times 
in removing low angular momentum gas in haloes with $M_{200}<0.7\times 10^{12}M_{\odot}$ and too weak
in haloes with $M_{200}>1.6\times 10^{12}M_{\odot}$.

\subsection{The baryonic Tully-Fisher relation}

\begin{figure}
\centering
\hspace{-0.7cm}\includegraphics[width=9cm]{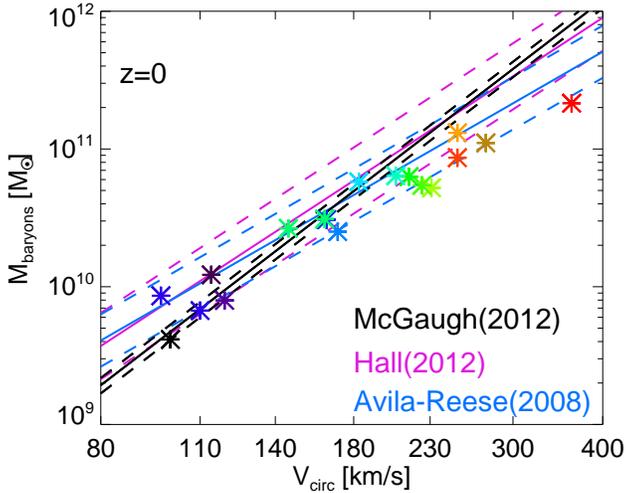}
\caption{Baryonic mass (within $r=20$ kpc) vs. circular velocity $V_{\rm circ}=\sqrt{GM(<r)/r}$ at $r=4R_{50}$ at $z=0$. 
Overplotted are the observational results from \citet{mcgaugh} (black), \citet{hall} (magenta) and \citet{avila} (cyan), 
for the so-called baryonic Tully-Fisher relation, with $1\sigma$ deviations represented by the dashed lines.}
\label{bartf}
\end{figure}

The luminosities and rotation velocities of disk galaxies are strongly correlated, they lie on the Tully-Fisher relation \citep{tully}.
In recent years, it has been shown that using the total baryonic mass instead of luminosity yields a tighter correlation \citep{mcgaugh}.
As McGaugh gives a relation that is among the steepest and tightest found in the literature, we also consider the observational
results of \citet{avila} and \citet{hall}, who find shallower slopes and significantly more scatter.
We attempt to compare to these observational relations in Figure \ref{bartf}. We plot the mass of stars and gas within
$r_{\rm gal}$ (as defined in Section \ref{rgal})
against the circular velocity $V_{\rm circ}=\sqrt{GM(<r)/r}$ at $r=4\; R_{50}$. We use this radius, as \citet{mcgaugh} use
measurements of velocities where rotation curves are flat, which is well reproduced 
by our choice as depicted in Figure \ref{vel}. Note that \citet{avila} use the maximum rotation velocity and \citet{hall} use
the velocity at half light radius. Only for the most massive, compact galaxies in our sample
which overproduce stars at low $z$, would a different choice of radius alter the data-points significantly.
A different choice for the radius within which we consider the baryonic mass does also not change our conclusions.

The baryonic Tully-Fisher relation is intrinsically connected to the $M_{\star}-M_{\rm halo}$ relation discussed in Figure \ref{moster} and
to the gas fractions discussed in Figure \ref{gasf}. Moreover as was shown e.g. in CS12, the correct distribution of matter
and thus the sizes of galaxies as discussed in Figure \ref{sizes} also play an important role. The Tully-Fisher relation is thus a good test for the
interplay of the model ingredients included in our galaxy formation code. Considering that for all the mentioned Figures we found at least
reasonable agreement with observation, with smaller deviations we expect also to find this in Figure \ref{bartf}. 
Compared to \citet{mcgaugh}, this is not, however, the case. Only a minority of galaxies lies within $1\sigma$ 
of  these observations and the power-law slope constituted by our model sample is closer to 3 as advocated
by \citet{avila} and not 4 as found by \citet{mcgaugh}. Due to the greater scatter found by \citet{hall}, our galaxies are marginally consistent
with their results and there is good agreement with the results of \citet{avila}.

Considering that the power-law slope of the relation given by our model galaxies is smaller
than most observations, it is interesting to discuss the trends.
The most massive galaxies have too high rotation velocities for their masses. This agrees with the fact that
they are too compact for their baryonic mass. There is a weaker trend for low-mass galaxies to lie above the relation,
possibly connected to too efficient feedback for these haloes preventing a further concentration of the halo.

In summary, our model galaxy sample shows a power-law slope for the baryonic Tully-Fisher relation, that agrees only with the lowest
in the range of values found observationally. The deviations from the relation can be connected to the
remaining fine-tuning problems for the feedback mechanisms which have already been discussed in the previous sections.

\subsection{Metals}

\begin{figure}
\centering
\hspace{-0.7cm}\includegraphics[width=9cm]{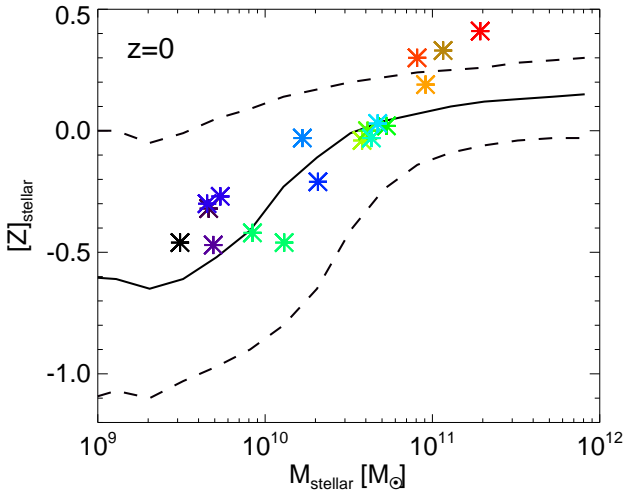}
\caption{Mean mass-weighted galactic stellar metallicity within $r=3\;R_{50}$, $[Z]$ vs. 
stellar galaxy mass $M_{\rm stellar}$ at $z=0$ compared to the observational results of
\citet{gallazzi}}
\label{z-stars}
\end{figure}

Another way to test how well the complex interplay of gas accretion, star formation, galactic winds and gas recycling
is modeled, is to consider the amount of metals stored in the galactic stars and gas. Metals are produced in stars
and returned to the ISM in stellar explosions or winds. Thus metal production and energetic feedback are intrinsically coupled.
Consequently, galactic winds are an efficient way to get metals out of galaxies. However, the metals in outflows can diffuse
into hot halo gas and re-accrete, alternatively, outflowing material can return in galactic fountains, providing ways for ejected
metals to re-enter the galaxy. Moreover, the metal content of galactic gas can be diluted by the accretion of pristine gas.
Basically all ingredients in our galaxy formation simulations have a direct influence on the galactic metallicities.

Whereas gas metallicities probe cumulative effects over the whole evolution of galaxies, stellar metallicities contain
probes from every stage in galaxy formation. In Figure \ref{z-stars} we attempt a comparison of the mean stellar
metallicities of our model galaxies to the observational data of \citet{gallazzi}. Apart from the too-metal-rich massive galaxies,
the sample agrees well with observations and reproduces the slope of total stellar metallicity $[Z]$ with stellar mass.

\begin{figure}
\centering
\hspace{-0.7cm}\includegraphics[width=9cm]{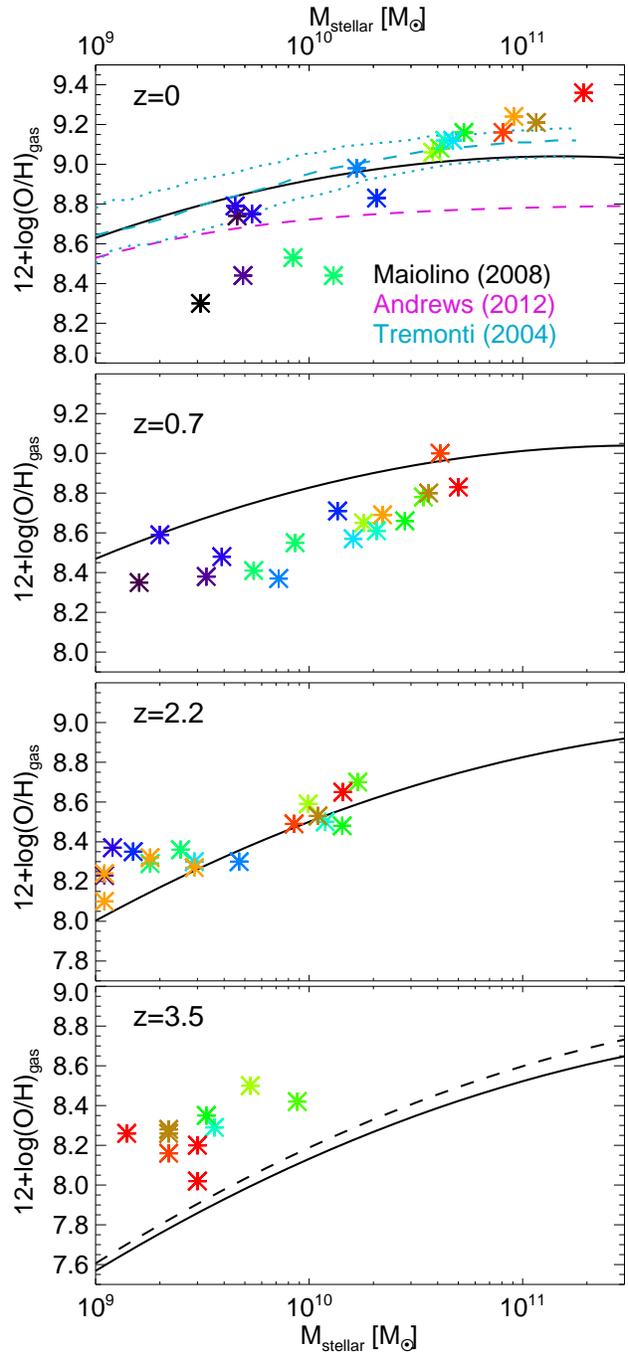}
\caption{Mean galactic gas-phase oxygen abundance $12 + log(O/H)$ vs. stellar galaxy mass at compared to the observational results of
\citet{maiolino}(black), \citet{andrews}(magenta) and \citet{tremonti} (cyan, dotted lines mark $1\sigma$ regions)
 at $z=0,0.7,2.2$ and $3.5$. At $z=3.5$, \citet{maiolino} give two calibrations.}
\label{ox-gas}
\end{figure}

The picture is different however for gas phase metallicities, which we explore in Figure \ref{ox-gas}. The observed relations are quite uncertain in their
calibration, which is why different authors find deviations of up to 0.3 dex in the normalization of the relation
of oxygen abundance $12+log(O/H)$ versus galaxy stellar mass. The slope also differs, but too a smaller amount.
Typically, quoted $1\sigma$ errors at $z=0$ are of the order 0.1-0.2 dex depending on mass and calibration.
We include three different observed $z=0$ relations by \citet{tremonti}, \citet{andrews} and by \citet{maiolino} and compare our results
to them. Our galaxies follow a significantly steeper relation with higher mass galaxies ($M_{\star}>2\times10^{10}M_{\odot}$)
agreeing with \citet{maiolino} and \citet{tremonti} and the lower mass galaxies in better agreement with \citet{andrews},
but on average more metal-poor than all three observational relations.

At higher redshifts, we do not find such a disagreement with the observational slope of the relation.
When we compare to the results of \citet{maiolino} at $z=0.7-3.5$, we find reasonable agreement between our models
and observations when taking into account the observational uncertainties. At $z=3.5$, our model galaxies are slightly too metal
rich, which considering the over-production of stars at $z>4$ is not surprising. At $z=2$ we find perfect
agreement with observations, whereas at $z=0.7$ our models are slightly metal-deficient, probably due to the 
small deficit in SFRs around $z=1$. The discrepancies at $z=0$ obviously origin at $z<0.7$.

At late times, four lower mass galaxies are metal deficient. They include two of the haloes with the lowest masses
and the two galaxies involved in the major merger of halo 0977. We have noted that these
galaxies are too gas rich. From observed relations, we know, that higher gas fractions and lower metallicities
are present in lower mass galaxies. As our lower mass galaxies are deficient in SF and too gas rich, late low metallicity infalling
material can dilute the metallicity and lead to even lower metallicities than the observed ones.
The most massive galaxies tend to be too metal-rich, which, as discussed for stellar metallicities,
can be connected to overly high low-$z$ SFRs.

\begin{figure}
\centering
\hspace{-0.4cm}\includegraphics[width=9cm]{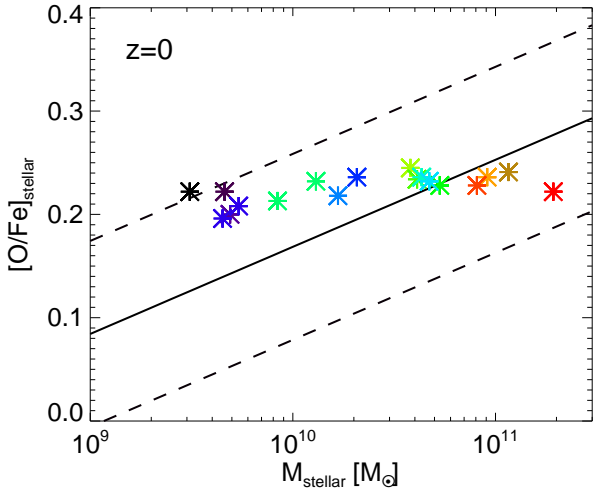}
\caption{Mean luminosity-weighted galactic stellar oxygen to iron ratio $[O/Fe]$ vs stellar galaxy mass at $z=0$
compared to the observational results of \citet{johansson} for elliptical galaxies.}
\label{ofe}
\end{figure}

It is well known that SNII and SNIa produce different fractions of elements and explode on different timescales. Because of that,
fractions of $O$ (SNII) to $Fe$ (SNIa) test the validity of the modeling of metal production yields,
SFHs and model SN time distributions. In Figure \ref{ofe}, we plot the mean luminosity-weighted element ratio $[O/Fe]$ in stars.
We compare them to observations from \citet{johansson} for elliptical galaxies. Although the average $[O/Fe]$ in our galaxies
agrees with ellipticals, these values are too high, as  disk galaxies typically show lower values of $[O/Fe]$ due to the higher
SFRs at late times. Compared to observed ellipticals, the increase in $[O/Fe]$ with galaxy mass is much shallower in our model
galaxy sample. Clearly, higher low-$z$ SFRs in lower mass galaxies would shift $[O/Fe]$ to lower values and the opposite is true
for the highest mass galaxies which overproduce stars at low redshifts. However, to shift the normalization, a different calibration
of yields and SN distributions over time is necessary. As we noted in Section \ref{metals}, the metal production yields we apply
were not calibrated in any way.

The modeling of turbulent diffusion of metals also has an influence on the metallicity in galaxies. We find
that using higher normalizations for the diffusion coefficients yields higher metallicities in galaxies.
Using the normalization proposed by \citet{greif} (see Section \ref{metdif}) increases the metallicity of model
galaxies by $\sim 0.2-0.4$ dex and would lead to significantly worse results compared to observations.
The reason for this is, that outflowing gas is metal-enriched and thus loses metals to the circumgalactic
gas by diffusion. The halo gas can later accrete onto the galaxy \citep{shen}. This effect is enhanced by stronger diffusion
coefficients.

Metals in galaxies are not equally distributed. Most galaxies show gas oxygen metallicities that decrease outwards,
with gradient slopes in units of dex per scale length being independent of galaxy mass \citep{zaritsky}.
As more massive galaxies are more extended, absolute gradients in dex/kpc are thus steeper for galaxies with 
lower masses and smaller sizes. Disk galaxies have steeper gradients than ellipticals and than interacting disks \citep{kewley}. 
\citet{zaritsky} found gradients ranging from -0.23 to +0.02 dex/kpc for a sample of $\sim 40$ disk galaxies with a mean
$\sim -0.06$ dex/kpc. Gradients are believed to arise from the inside-out formation of disk galaxies and the corresponding
variation of SF and feedback efficiency with radius.

\begin{figure}
\centering
\hspace{-0.4cm}\includegraphics[width=9cm]{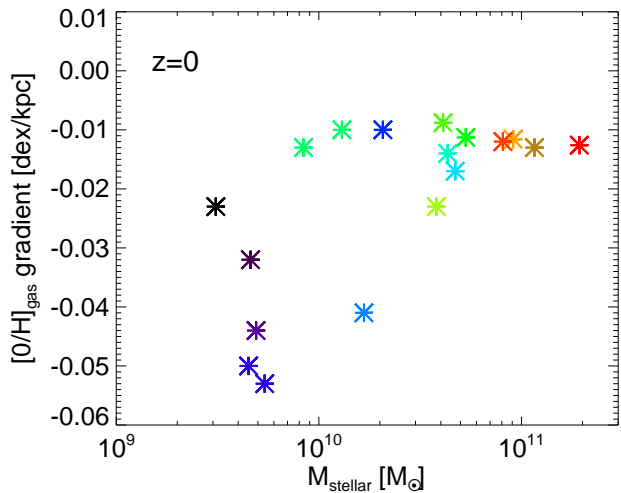}
\caption{The (lack of) gas phase metallicity gradients in the galactic disks at $z=0$. Mean gradients in $[O/H]$ are plotted
against galactic stellar mass $M_{\rm stellar}$.}
\label{gradients}
\end{figure}

In Figure \ref{gradients}, we plot the gradients which we find in the oxygen abundances in the gas disks of our model galaxies and plot
them against galaxy mass. All our galaxies show clear negative gradients, but they are rather shallow ranging from -0.053 to -0.007 dex/kpc
, being a factor of $\sim 3-4$ shallower than observations. We do however reproduce the fact that absolute gradients are
steeper for smaller galaxies. As far as our interacting galaxies in the merger haloes 2283 and 0977  are concerned, the extended disks in halo 0977
indeed have flatter gradients than the other galaxies of similar mass. The significantly more compact galaxies in halo 2283 however show
the steepest gradients in our sample.

It is interesting that \citet{pilkington} found steeper gradients in a set of cosmological simulations with AMR and SPH codes.
Considering our modeling there are two ingredients which can make gradients become too flat. Turbulent diffusion of metals is modeled on
the scale of the SPH smoothing kernel and thus is not independent of resolution. In test simulations, we find that stronger diffusion produces flatter
metallicity gradients, as diffusion intrinsically acts against metallicity gradients. 
Moreover, our two phase enrichment model, which spreads half of the metals into the hot gas phase,
intrinsically yields a relatively non-local enrichment of the ISM, which also weakens gradients.

Recently, \citet{Gibson} compared gradients in cosmological simulations which share initial conditions but use different
feedback prescriptions. They found that models which assume efficient SN feedback and pre-SN early stellar feedback produce
shallower metallicity gradients than more conservative feedback models, as they spread metals over larger scales. As our models, 
their stronger feedback models also do not predict a steepening of gradients at high-$z$, a behaviour displayed by their weaker
feedback models. In general, their conclusions about the interplay between feedback and gradients thus agree well with ours.

In summary, our model galaxies reproduce the observed relation of stellar metallicity vs. stellar mass well.
The evolution of the gas phase oxygen abundance is also reproduced reasonably well at $z\ge1$. At low-$z$ the problems
with SFHs discussed in Section \ref{rgal} lead to deviations from observation. The oxygen-to-iron fractions [O/Fe] in 
our models are too high and the metallicity gradients in our gas disks are too shallow. This indicates that the details
of our modeling of metal production, enrichment and diffusion still needs fine-tuning.

\section{Conclusions}

We have presented an extensive update to the multiphase SPH galaxy formation code by \citet{cs05,cs06}. 
This includes updated metal cooling rates based on individual element abundances, a continuous modeling
of metal production in SNIa and AGB stars, an increase in the star formation threshold and a model
for the turbulent diffusion of gas phase metals. Moreover, we now use feedback prescriptions which spread
energy in kinetic and thermal form and apply a model for the effect of radiation pressure from 
massive, young stars on the surrounding ISM.

We calibrate our feedback model to match the SFH of the previously well studied halo Aquarius C (CS09, CS12)
predicted by results from abundance matching techniques out to $z=4$ (MNW13).
By doing so, we confirm the findings of \citet{stinson13} that modeling feedback from young stars can play
a very important role in shifting star formation to later times, in increasing the efficiency of feedback at high $z$ and 
thus in improving cosmological simulations of galaxy formation. We find that in order to reproduce the inferred
SFH, the influence of radiation pressure from young stars on the ISM has to be stronger at higher redshifts.
We achieve this by invoking a model in which the higher gas velocity dispersions at early formation stages lead to
larger star formation regions and thus to greater column densities of dust, which enhance the effect of radiation pressure.
Our model still predicts an overproduction of stars at $z>4$, however, pointing to inaccurate modeling of SF at early formation stages.

It was noted in CS12 that there is a priori no reason why
this halo should not be an outlier in terms of its SFH.
However, considering the many previous galaxy formation models for halo AqC from CS12, our model not only improves
the shape of the SFH, but also yields the most dominant disk galaxy in this halo known to us. The galaxy properties show
good agreement with a variety of observations. 

The efficiency of the coupling between the radiation field and the dusty gas is uncertain and some have argued that
the approach of \citet{hopkins}, which is similar to our modeling, overestimates the effects of radiation pressure \citep{krumholz}.
The influence of our radiation pressure model on the SFH is, however, a key ingredient for improving the 
agreement of simulated galaxies with observations. We conclude that any kind of feedback process acting before the explosion
of SNe with effects similar to those in our model is likely to achieve this effect (see also \citealp{stinson13,4disk}). 

We have applied our code to a sample of 16 haloes with virial masses $M_{200}$ ranging from $1.7\times10^{11}M_{\odot}$ to
$2.4\times10^{12}M_{\odot}$, including haloes which were selected to have no recent major mergers \citep{aquarius} 
and haloes with low-$z$ mergers \citep{oser}.
As a population the resulting galaxies reproduce the observed $M_{\star}-M_{\rm halo}$-relation well at all redshifts
$z=0-4$. There are however trends of too high stellar masses at high $z$, of low-mass
haloes under-producing stars at low $z$, and of high-mass haloes overproducing stars at late times. 

Our model galaxies also reproduce observations of gas fractions, stellar half-mass radii, star formation rates
and gas phase metallicities at $z=0-3$ reasonably well.  At $z=0$, our sample also shows agreement with
observed stellar metallicities and the galaxies provide a reasonable fit to the baryonic Tully-Fisher relation.
Deviations from the observed scaling relations
can mostly be connected to the problems with SFHs mentioned above.
Considering these results, one could argue that our finding that reproducing the SFH
helps in reproducing a range of observations to some degree validates the abundance matching methods used by MNW13.

The problems at the high mass end could well be explained by the lack of a model for AGN feedback,
which could prevent the cooling of large masses of halo gas at low $z$, which occurs in our models (see e.g. \citealp{puchwein}).
The low-mass haloes indicate that our model is far from perfect, as feedback in these galaxies 
is too efficient at preventing star formation at late times. Again, the question arises whether our calibration method is appropriate.
As the peak in conversion efficiency $M_{\star}/M_{\rm halo}$ appears to happen at slightly lower masses than that of AqC, it is justified
to note that lower halo masses could be more appropriate for calibration. \citet{4disk} argue that calibrating their
galaxy formation model at a halo mass $M_{200}<10^{11.6}M_{\odot}$ enables them to produce galaxies that reproduce a range
of observations up to the peak in conversion efficiency.

Compared to simulations with the previous code version (CS09) the resulting morphologies are significantly more
disk-dominated. 15 out of 16 models show stellar disk components.  Morphologies include
spheroidal components, extended gas disks, disks that are disturbed and thickened by mergers,
warped gas and stellar disks, disks suffering from reorientation and misaligned infall, barred disks,
a galaxy with counter-rotating disks living on top of each other, thin extended
disks with varying bulge fractions and massive disks, which are compact and thick.
Radial surface brightness profiles of our galaxies show pure exponentials and bulge + exponential disk
profiles, as well as up- and down-bending breaks,
as found in observations \citep{s4g}. Moreover, apart form two massive galaxies, all circular velocity curves are flat
and none show strong central peaks of the kind encountered in many previous simulations (e.g. some models in CS12).

We show how distributions of stars over the circularity $\epsilon$ vs. formation time $a_{\rm form}$ plane can be used to visualize formation histories
and extract disk populations. This tool nicely reveals mergers and misaligned infall of gas. Interestingly, if misaligned
infall and/or the reorientation of existing disks occur, the signatures of these processes which heat stellar populations,
are very similar to those observed for idealized, semi-cosmological models studied by AW13. 
These processes are of interest in the context of the formation of counter-rotating components and $2\sigma$ signatures
found in several elliptical galaxies \citep{sauron}.

Most of the disks in our model sample feature a well-defined start for disk formation, defined by a merger or a misaligned infall
event. The later the last destructive event occurs, the lower the disk mass. These 'disk ages' all correspond to $z<2$. It is 
interesting that the age inferred for the thin disk population in the solar neighbourhood is even older (see \citealp{ab} and
references therein). It should also be noted that observational evidence in favour of a picture where a significant fraction
disk galaxies (re)-formed after a destructive event since $z\sim2$ exists \citep{puech}, consistent with the idea that the Milky Way's
formation history was more quiescent than that of an average disk galaxy \citep{hammer}.

The disk fractions derived from the $\epsilon-a_{\rm form}$ plots range from 15 to 65 \%. Some galaxies are thus disk dominated. 
However, there are real galaxies with even higher disk fractions.  Considering the shortcomings of our SFHs at $z>4$, and assuming 
disk formation times are correctly modeled, we can estimate disk fractions for our haloes under the assumption that SFHs
follow the MNW13 predictions. For our three most disk-dominated galaxies, these estimations yield 
an increase in disk fraction from 59-65 \% to 70-77 \%. We thus conclude
that high disk fractions are not problematic for $\Lambda$CDM haloes with sufficiently quiet merger histories.
The Aquarius haloes show the highest disk fractions and the quietest merger histories, as was expected from their selection.
However, previous simulations (CS09, CS12) failed to validate these predictions.

Some of the problems of our simulations are connected to metallicities. Although our model galaxies reproduce
the stellar metallicity - stellar mass relation, the $z=0$ gas phase metallicity - stellar mass relation is too steep,
our alpha-element abundances are too high and the metallicity gradients in our disks are too shallow.
No optimization of yields has been attempted for our models so far, leaving room for improvement.
Shallow metallicity gradients are likely to be connected to the details of our diffusion model and our two-phase enrichment
scheme, which both could require modification. Another problem is that tests indicate unsatisfactory convergence
of our results when increasing the resolution. Such problems are common for galaxy formation codes (see e.g. the discussion in CS12).
One must take into account that important ingredients are 'sub-grid' and thus that a dependence on the lowest resolved
length-scale is expected when resolution is increased without adjusting model parameters.

In general, our simulations provide us with an interesting sample of simulated galaxies with realistic properties at least for MW-like halo masses
$M_{200}\sim 10^{12}M_{\odot}$. The simulations predict many more observables that can be compared to the real galaxy population
than are touched in this paper. 

\section*{Acknowledgments}

We thank the referee, Brad Gibson, for valuable comments.
MA acknowledges support from the DFG Excellence Cluster "Origin and Structure of the Universe".
SW was supported in part by Advanced Grant 246797 "GALFORMOD" from the European Research Council.

\end{document}